\documentclass[usenatbib,preprint]{mn2e}

\usepackage{graphicx}
\usepackage {natbib,aas_macros}
\usepackage {bm}
\usepackage {amsmath,amssymb}
\usepackage {setspace}

\newcommand{\cm}{~{\rm g~cm}^{-3} } 
\newcommand{\magB}{\mathbf{B}}
\newcommand{\ele}{\mathbf{E}}
\newcommand{\cul}{\mathbf{J}}
\newcommand{\vel}{\mathbf{v}}

\title[Effects of Ohmic and ambipolar diffusion on formation of circumstellar discs]{Effects of Ohmic and ambipolar diffusion on formation and evolution of first cores, protostars and circumstellar discs}

\author[Tsukamoto et al]{
Y. Tsukamoto$^{1,2}$,  K. Iwasaki$^{2,3}$, S. Okuzumi$^{4}$, M. N. Machida$^{4}$, and  S. Inutsuka$^{2}$ \\
$^1$Laboratory of Computational Astrophysics, RIKEN, Saitama, Japan  \\
$^2$Department of Physics, Nagoya University, Aichi, Japan  \\
$^3$Department of Environmental Systems Science,
Faculty of Science and Engineering, Doshisha University, Kyoto, Japan \\
$^4$Department of Earth and Planetary Sciences, Tokyo Institute of Technology, Tokyo, Japan \\
$^5$Department of Earth and Planetary Sciences, Kyushu University, Fukuoka, Japan \\
}

\begin{document}
\maketitle

\begin{abstract}
We investigate the formation and evolution of a first core, protostar, and 
circumstellar disc with a three-dimensional non-ideal (including both Ohmic 
and ambipolar diffusion) radiation magnetohydrodynamics simulation.
We found that the magnetic flux is largely removed by magnetic diffusion 
in the first core phase and that the plasma $\beta$ of the centre 
of the first core becomes large, $\beta>10^4$.
Thus, proper treatment of first core phase is crucial in
investigating the formation of protostar and disc.
On the other hand, in an ideal simulation, $\beta\sim 10$ 
at the centre of the first core.
The simulations with magnetic diffusion show 
that the circumstellar disc forms at almost the same time 
of protostar formation even with a relatively strong initial magnetic 
field (the value for the initial mass-to-flux ratio of 
the cloud core relative to the critical value is $\mu=4$).
The disc has a radius of $r \sim 1$ AU at the protostar formation epoch.
We confirm that the disc is rotationally supported.
We also show that the disc is massive ($Q\sim 1$) and that
gravitational instability may play an important 
role in the subsequent disc evolution.
\end{abstract}


\section{Introduction}
\label{intro}
The molecular cloud core is the formation site of the star.
Already almost half-a-century ago, \citet{1969MNRAS.145..271L} investigated 
the formation process of the protostar 
with one-dimensional radiation hydrodynamics simulation starting 
from a gravitationally unstable cloud core.
An overview of the evolution obtained from that simulation is as follows:
While the dust thermal emission
effectively removes the thermal energy generated 
by the compressional heating caused by the gravitational collapse,
the gas evolves almost isothermally.
At $ \rho  \sim 10^{-13}{\rm g ~cm^{-3}}$, the compressional heating 
overtakes the radiative cooling and the gas begins to evolve 
adiabatically.
In this adiabatic evolution phase, the temperature 
evolves as $T \propto \rho^{\gamma-1}$, where $\gamma$ is the 
adiabatic index ($\gamma=5/3$ for $T \lesssim 100$ K 
and $\gamma=7/5$ for $100 \lesssim T \lesssim 2000$ K).
Because this index is larger than the critical adiabatic index 
for spherical gravitational collapse, $\gamma_{\rm crit}=4/3$, the gravitational 
collapse temporarily halts and a quasi-hydrostatic core forms, commonly 
known as the {\it first core}.
When the central temperature of the first core 
reaches $\sim2000$ K, the hydrogen molecules begin to dissociate.
This endothermic reaction changes the effective 
adiabatic index to $\gamma_{\rm eff}=1.1$.
Because this is smaller than $\gamma_{\rm crit}$,
the gravitational collapse resumes, which is 
known as the  {\it second collapse}.
Finally, when the molecular hydrogen is completely dissociated, 
the gas evolves adiabatically again and the gravitational collapse finishes.
The adiabatic core formed at the centre is the protostar (or the second core).
This evolution process was later confirmed, more 
sophisticated one-dimensional 
simulations \citep{2000ApJ...531..350M,2012A&A...543A..60V,2013A&A...557A..90V}.

Although, the general picture of the formation process
of the protostar was established by \citet{1969MNRAS.145..271L}
with one-dimensional simulations, multidimensional simulations
are necessary to investigate important phenomena such as
the formation and evolution of the circumstellar disc.
After the radiation hydrodynamics simulations 
done by \citet{1969MNRAS.145..271L}, it took 
several decades to develop and perform three-dimensional 
radiation hydrodynamics
simulations of gravitational collapse 
\citep{2006MNRAS.367...32W,2010MNRAS.404L..79B,
2011MNRAS.417.2036B,2013ApJ...763....6T,2013MNRAS.436.1667T,2015MNRAS.446.1175T}.
These studies revealed that the multi-dimensionality causes 
new and interesting phenomena.
For example, \citet{2010MNRAS.404L..79B} found that the bipolar 
outflow from the first core can be driven by radiative 
feedback from the protostar.
\citet{2015MNRAS.446.1175T} investigated the evolution 
of the circumstellar discs in the unmagnetized cloud core 
and found that the temperature structure of the disc
is determined by diffusive radiative transfer in the radial direction 
in its early evolution phase.

The magnetic field is another important ingredient 
in the star formation process.
Observations suggest that the molecular cloud cores are 
magnetized \citep[e.g.][]{2005ApJ...624..773H,2008ApJ...680..457T}.
\citet{2008ApJ...680..457T} showed that the mean value of the mass-to-flux 
ratio relative to the critical value, $\mu$, of the nearby dark cloud cores 
is $\mu \sim 2-3$ and suggested that the magnetic field of the typical cloud 
core is relatively strong.
The magnetic field drives the outflow from both 
the first core and the protostar.
The typical velocity of the outflow is determined by the rotational 
velocity at the launching point 
($v\sim 2$ km/s from the first core and $v\sim 20$ km/s 
from the protostar) \citep{2002ApJ...575..306T,
2008ApJ...676.1088M,2008A&A...477....9H,2012MNRAS.423L..45P}.
Another important effect caused by the magnetic field is the removal of
the gas angular momentum. This effect is known 
as {\it magnetic braking} \citep[][]{1979ApJ...230..204M}.
Until recently, it was believed that the disc formation is a 
natural consequence of the gravitational collapse of 
a rotating molecular cloud core.
Actually, three-dimensional simulations, with a weak 
magnetic field or without it, show 
that a relatively large circumstellar disc 
(with a radius of several tens of AU) 
develops in the early phase of protostar 
formation \citep{1998ApJ...508L..95B,2011MNRAS.417.2036B,2011MNRAS.416..591T, 2013MNRAS.428.1321T,2013MNRAS.436.1667T}
However, previous works with ideal magnetohydrodynamics (MHD) simulations have
shown that the relatively strong magnetic 
field ($\mu\sim 1$) completely suppresses 
the formation of a rotationally supported disc around the protostar 
at its formation epoch \citep{2008ApJ...681.1356M,2008A&A...477....9H}.

Ideal MHD is, however, not a good approximation for
the simulations of the magnetized molecular cloud core. 
Because the ionization degree of the 
cloud core is quite low, it is expected that 
non-ideal magnetic effects such as Ohmic diffusion, 
Hall effect, and ambipolar diffusion play important roles 
during the formation and evolution of the circumstellar disc.

The influence of non-ideal magnetic effects on the 
disc formation is still controversial.
\citet{2011ApJ...738..180L} investigated the influences of 
the non-ideal magnetic effects.
They pointed out that ambipolar diffusion is 
the dominant diffusion process of the 
magnetic field and concluded that neither Ohmic nor ambipolar diffusion weakens 
the magnetic braking and that the disc formation is still strongly suppressed 
even with the magnetic diffusion.
On the other hand, \citet{2011PASJ...63..555M} showed 
that a relatively large disc of about a few tens of AU in size 
forms in the early phase of the protostar formation 
although they considered only Ohmic diffusion.

The discrepancy could come from the difference in the initial conditions 
and the treatment of the inner boundary (or a sink at the centre) 
of the simulations.
In the simulations of \citet{2008ApJ...681.1356M} 
and \citet{2011ApJ...738..180L}, 
the inner boundary or sink is set from the beginning of the simulations.
In such a set-up, the simulations cannot follow the evolution of a first core
which is mainly supported by gas pressure and 
not necessarily by rotation.
Although the first core is a transient object, its density is high enough 
that the magnetic flux is efficiently removed from the first core during its 
evolution \citep{2012A&A...541A..35D}.
Furthermore, it is suggested that the greater part of the
first core directly becomes the circumstellar 
disc \citep{2011MNRAS.413.2767M} just after the protostar formation.
Therefore, calculating the first-core phase correctly in the simulations 
is crucial to investigate the very early phase of disc evolution.
On the other hand, \citet{2011PASJ...63..555M} used sink cells
with ``threshold density". In their simulations, the sink cell takes in
the gas when its density becomes larger than the threshold density.
In this case, the gas whose density is smaller than the threshold
density can reside inside or around the sink cell 
regardless of whether the gas is rotationally supported or not.
This treatment may also affect the disc evolution process.
\citet{2014MNRAS.438.2278M} showed that 
the sink treatment (its radius and 
the threshold density) significantly
affects the formation and evolution of the circumstellar disc.

To reveal the realistic formation 
and evolution processes of the first core, the protostar, 
and the circumstellar disc,
appropriate treatment of the radiation transfer in the simulation is crucial,
because the magnetic diffusion coefficients are functions of temperature.
The previous studies with MHD simulations mentioned above do 
not include radiation transfer and employ a simplified equation of 
state (EOS) which mimics the temperature evolution 
of the {\it centre} of the cloud core.
We call this the barotropic approximation.
The simulations with radiation transfer, however, have shown 
that the temperature structures in the first core or around the 
protostar are strikingly different from those expected from
the  barotropic approximation 
\citep{2006MNRAS.367...32W,2010MNRAS.404L..79B,
2013ApJ...763....6T,2015MNRAS.446.1175T}.

Three-dimensional simulations which include 
both the magnetic field and radiation transfer have 
not been successful until recently.
\citet{2013ApJ...763....6T} was the first to succeed 
with such a simulation with a 
grid code and found that the Ohmic diffusion alters the structure 
around the protostar significantly.
With ideal radiation magnetohydrodynamics (RMHD) simulations using the 
smoothed particle hydrodynamics (SPH) method, \citet{2014MNRAS.437...77B} 
also investigated the formation and evolution of the protostar, 
especially the long-term evolution of 
the bipolar jets driven around the protostar.
They showed that the jets heat up the gas in the envelope after they 
break up the remnant of the first core.
Such a radiative heating process may affect the ionization degree of the 
gas and change the magnetic diffusion coefficients.
However, \citet{2014MNRAS.437...77B} did not consider magnetic 
diffusion processes.

As pointed out in previous studies \citep{2011ApJ...738..180L}, it is 
expected that ambipolar diffusion will play a role during 
the formation process of the protostar and the disc around it.
Very recently, \citet{2015ApJ...801..117T} conducted a simulation with 
both Ohmic and ambipolar diffusion.
However, they only calculated the evolution until the end 
of the first core phase with ambipolar diffusion and the effect of the 
ambipolar diffusion is still unclear.


In this paper, we investigate the formation of the first core, protostar, and 
the circumstellar disc using a three-dimensional non-ideal RMHD simulation.
We employ the SPH method and use it to produce the first results of 
the three-dimensional non-ideal RMHD simulations with SPH.
Here, we focus on the effects of magnetic (Ohmic and ambipolar) diffusion, but 
do not include the Hall effect.
To avoid the numerical artefact caused by the sink, we do not 
introduce it, but rather investigate the structure around the 
protostar to determine whether the formation of the circumstellar disc is 
possible at the very early phase of protostar formation.
This paper is organized as follows:
In \S 2, we briefly describe the non-ideal magnetohydrodynamic effects.
In \S 3, we describe the numerical method and initial conditions 
for the simulations, the results of which are given in \S 4, 
and then summarized and discussed in \S5.

\section{Non-ideal  magnetohydrodynamic effects}
The ionization degree in the molecular cloud core is quite low and
the gas can be regarded as weakly ionized plasma.
In weakly ionized plasma, the microscopic collisions 
between neutral, positively-charged, and negatively-charged particles
produce finite conductivity and 
non-ideal magnetohydrodynamic effects, or in short, non-ideal effects arise.

The non-ideal effects appear as the correction terms 
in the induction equation. They can be derived by calculating
the drift velocity of the charged particles.
Here, we derive the induction equation for the weakly ionized plasma 
according to \citet{1999MNRAS.303..239W} and \citet{2007Ap&SS.311...35W}.

We start with
\begin{eqnarray}
\label{induct}
\frac{\partial \magB}{\partial t}=- c \nabla \times \ele,\\
\cul= \frac{c}{4\pi} \nabla \times \magB.
\end{eqnarray}
where $\magB$ is the magnetic field,
$\cul$ is the current density,
$\ele$ is the electric field, and
$c$ is the speed of light.
By the Lorentz transformation to the rest frame of the fluid (that is
essentially the rest frame of bulk of neutral particles), 
the electric field becomes
\begin{eqnarray}
\label{lorentz_tra}
\ele'=\ele+\frac{\vel \times \magB}{c}.
\end{eqnarray}
Here, $\vel$ and $\ele'$ are the fluid velocity 
and the electric field in the rest frame of the fluid, respectively.
The conductivity in the weakly ionized plasma can be calculated using
the balance of the force that acts on the charged particles,
\begin{eqnarray}
\label{force_balance}
Z_je(\ele'+\frac{\vel_j \times \magB}{c})-\gamma_j \rho m_j \vel_j=0.
\end{eqnarray}
Here, subscript $j$ denotes the species of charged particles,
$Z_je$  is the charge,
$\vel_j$ is the relative velocity of charged particles in the fluid rest frame,
$\gamma_j=\langle \sigma v \rangle_j/(m_j+m)$ 
where $\langle \sigma v \rangle_j$ 
is the rate coefficient for momentum transfer,
$m_j$ is the mass of charged particles, $m$ is the mean mass of 
neutral particles, and
$\rho$ is the density of neutral particles.
Note that, in the weakly ionized plasma, most of the particles are neutral 
and the inertia of the charged particles and the collisions with other 
charged particles are negligible.
Note also that, under the MHD approximation, 
the difference between the magnetic field and the current density 
in computation frame and those in the rest frame is negligible.
We assumed the local charge neutrality $\sum_j n_j Z_j =0$.
By inverting (\ref{force_balance}) for $\vel_j$ and calculating the
current density, $\cul=\sum_j n_j Z_j e \vel_j$, we obtain
\begin{eqnarray}
\label{generalized_ohm}
\cul=\sigma_O \ele'+\sigma_H {\hat \magB} \times \ele' - (\sigma_P-\sigma_O) {\hat \magB} \times {\hat \magB} \times \ele',
\end{eqnarray}
where
\begin{eqnarray}
\sigma_O&=&\frac{ec}{B}\sum_j n_jZ_j\beta_j,\\
\sigma_H&=&\frac{ec}{B}\sum_j\frac{n_jZ_j}{1+\beta_j^2},\\
\sigma_P&=&\frac{ec}{B}\sum_j\frac{n_j Z_j \beta_j}{1+\beta_j^2},\\
\end{eqnarray}
are the Ohmic, Hall, and Pedersen conductivities, respectively. 
Here, $\beta_j=Z_jeB/(m_jc\gamma_j\rho)$ is the Hall parameter which is the product of the cyclotron frequency 
and stopping time.
Finally, by inverting equation (\ref{generalized_ohm}) for $\ele'$
and using equation (\ref{induct}) and (\ref{lorentz_tra}),
we obtain
\begin{eqnarray}
\label{generalized_induction}
\frac{\partial \magB}{\partial t} &=& \nabla \times(\vel \times \magB) \\
&-& \nabla \times \left\{ \eta_O (\nabla \times \mathbf B) 
+\eta_H (\nabla \times \mathbf B)  \times \mathbf {\hat {B}}
- \eta_A ((\nabla \times \mathbf B) \times \mathbf {\hat {B}}) \times \mathbf {\hat {B}}\right\} \nonumber.
\end{eqnarray}
This is the induction equation with non-ideal effects.
The second, third, and fourth term on the right hand side of equation
(\ref{generalized_induction}) describe the Ohmic diffusion, Hall term, and
ambipolar diffusion, respectively.
Here,
\begin{eqnarray}
\eta_O&=&\frac{c^2}{4 \pi \sigma_O},\\
\eta_H&=&\frac{c^2\sigma_H}{4 \pi (\sigma_H^2+\sigma_P^2)},\\
\eta_A&=&\frac{c^2\sigma_P}{4 \pi (\sigma_H^2+\sigma_P^2)}-\eta_O,
\end{eqnarray}
are 
the Ohmic,
Hall, and
ambipolar diffusion coefficients, respectively.
In this paper, the Hall term is neglected owing to the numerical difficulty 
associated with it.
The effect of the Hall term will be investigated in future works.

We constructed the data table of the diffusion coefficients 
by calculating a chemical reaction network 
of ${\rm H_3^+,~HCO^+, Mg^+, He^+, C^+, H^+, e^-}$ in gas phase and 
the positively-charged, neutral, and negatively-charged dust grain of
uniform size using the methods 
described in \citet{2002ApJ...573..199N} and \citet{2009ApJ...698.1122O}. 
We assumed that the dust to gas ratio is $10^{-2}$. We also assumed 
that the dust grain size and density are $a=3.5 \times 10^{-2} ~{\rm \mu m}$ and $\rho_d=2 \cm$, respectively. 
We considered non-thermal ionization by the cosmic rays and thermal 
ionization in our calculations. The cosmic-ray ionization rate was fixed 
to be $\xi_{\rm CR}=10^{-17} s^{-1}$.
When the temperature reaches $T\sim 1000$ K, thermal ionization
is the dominant source of ionization. 
In this paper, we consider the effect of the thermal ionization
by considering the thermal ionization of potassium.
The coupling between the magnetic field and the gas quickly recovers around
$T\sim 1000$ K
because the thermal ionization provides a sufficient ionization degree.

\begin{figure}
\includegraphics[width=60mm,angle=-90]{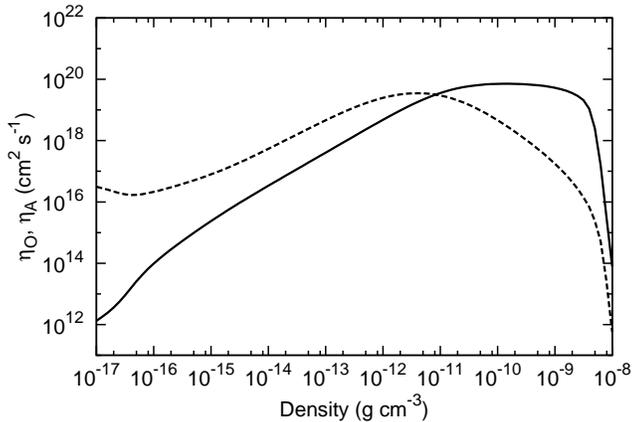}
\caption{
Diffusion coefficients, $\eta_O$ (solid) and $\eta_A$ (dotted) 
as a function of density.
For this plot, we assumed that the temperature and magnetic field are 
functions of density (see eq. (\ref{rho_dep})).
}
\label{eta_plot}
\end{figure}

In figure \ref{eta_plot}, we show the Ohmic and ambipolar 
diffusion coefficients under the typical evolution of the gas.
To make figure \ref{eta_plot}, we assumed that 
the temperature and magnetic field change as,
\begin{eqnarray}
\label{rho_dep}
B(\rho)=100 \left( \frac{\rho}{10^{-15} \cm}\right)^{2/3} ~{\rm \mu G} , \nonumber \\
T(\rho)=10 \left \{1+ \left(\frac{\rho}{10^{-13} \cm}\right)^{2/5} \right \} {\rm K}.
\end{eqnarray}
The figure shows that  
the diffusion coefficients suddenly drop
around $\rho = 5\times 10^{-9} \cm$ where the 
temperature is about $T = 1000$ K 
and the ionization degree quickly
increases owing to the thermal ionization of potassium.

\section{Numerical Method and Initial Conditions}
\label{method}
In this study, we solve the non-ideal radiation 
magnetohydrodynamics equations with self-gravity,
\begin{eqnarray}
\frac{D \mathbf{v}}{D t}&=&-\frac{1}{\rho}\left\{ \nabla
\left( P+\frac{1}{2}B^2 \right) - \nabla \cdot (\mathbf{ B B})\right\}  - \nabla \Phi,  \\
\frac{D}{D t}\left( \frac{\mathbf B}{\rho}\right) &=&\left(\frac{\mathbf B}{\rho} \cdot \nabla \right)\mathbf v   \\ 
&-& \frac{1}{\rho} \nabla \times \left\{ \eta_O (\nabla \times \mathbf B) - \eta_A (\nabla \times \mathbf B) \times \mathbf {\hat {B}} \times \mathbf {\hat {B}}\right\}, \nonumber \\
\frac{D}{D t} \left ( \frac{E_r}{\rho} \right ) &=& - \frac{\nabla \cdot \mathbf{F}_{\bm r}}{\rho} - \frac{\nabla \mathbf{v} : \mathbb
{P}_r}{\rho} +\kappa_P c (  a_r T_g^4 - E_r),  \\
\frac{D}{D t} \left ( \frac{e}{\rho} \right ) &=& - \frac{1}{\rho} \nabla \cdot \left \{ ( P+\frac{1}{2}B^2) \mathbf{v} -\mathbf{B} (\mathbf{B}\cdot\mathbf{v}) \right \} \nonumber \\
&-& \kappa_P c (  a_r T_g^4 - E_r)-\mathbf{v}\cdot\nabla \Phi  \nonumber \\
&-& \frac{1}{\rho} \nabla \cdot \left[ \left\{(\eta_O (\nabla \times \mathbf B) \right. \right. \nonumber \\
 &-&  \left. \left. \eta_A (\nabla \times \mathbf B) \times \mathbf {\hat {B}} \times \mathbf {\hat {B}}\right\} \times \mathbf B\right], \\
\nabla^2 \Phi&=&4 \pi G \rho.
\end{eqnarray}
Here, $\rho$ is the gas density, $\mathbf{v}$ is the velocity,
$~\mathbf{B}$ is the magnetic field, 
$~\mathbf{\hat{B}}$ is the unit directional vector of the magnetic field, 
$P$ is the gas pressure, 
$E_r$ is the radiation energy, 
$\mathbf{F}_{\bm r}$ is the radiation flux,
$\mathbb{P}_r$ is the radiation pressure,
$T_g$ is the gas temperature,
$\kappa_P$ is the Plank mean opacity, 
$e=\rho u+\frac{1}{2}(\rho \mathbf{v}^2+\mathbf{B}^2)$ is 
the total energy with $u$ specific internal energy,
 and $\Phi$ is the gravitational potential.
Parameters, $a_r$ and $G$ are the radiation and
gravitational constants, respectively.

 We adopt the gray approximation (frequency-integrated radiation
transfer) and we assume local thermodynamic equilibrium (LTE).
To close the equations for radiation transfer, 
we employ flux-limited diffusion (FLD) approximations,
\begin{eqnarray}
\mathbf{F}_{\bm r}&=&\frac{c\lambda}{\kappa_R \rho}\nabla E_r,\hspace{1em}
\lambda(R)=\frac{2+R}{6+2R+R^2},\nonumber\\
R&=&\frac{|\nabla E_r|}{\kappa_R \rho E_r}, \hspace{1em}
\mathbb{P}_r=\mathbb{D}E_r,\nonumber \\
\mathbb{D}&=&\frac{1-\chi}{2}\mathbb{I}+\frac{3\chi-1}{2}\mathbf{n}\otimes \mathbf{n},\hspace{1em}
\chi=\lambda+\lambda^2R^2,\hspace{1em} \nonumber \\
\mathbf{n}&=&\frac{\nabla E_r}{|\nabla E_r|}.\nonumber
\end{eqnarray}
Here, $\kappa_R$ is the Rosseland mean opacity.
FLD is a diffusion scheme which is designed to maintain the causality of 
$|\mathbf{F}_{\bm r}|<c E_r$. 
It is suitable for optically thick gas owing to its diffusive nature.
In this paper, we use the SPH method to investigate 
the formation of a protostar and disc.
The SPH method can be easily implemented and
is suitable for simulations which treat the large dynamic range 
because of its adaptive nature.
The ideal MHD part was solved by adopting the Godunov smoothed particle magnetohydrodynamics (GSPMHD) method 
in which the Godunov method and the method of characteristics are
used to calculate the interactions
between the particles instead of artificial dissipation terms
\citep{2011MNRAS.418.1668I}.
The divergence-free constraint on the magnetic field was maintained using
the hyperbolic divergence cleaning method for
GSPMHD \citep{2013ASPC..474..239I}.
The radiative transfer was treated by the FLD-SPH method 
\citep{2004MNRAS.353.1078W,2005MNRAS.364.1367W}.
We treated Ohmic and ambipolar diffusion with the method described by
\citet{2013MNRAS.434.2593T} and
\citet{2014MNRAS.444.1104W}, respectively.
Both diffusion processes were accelerated by super time stepping method 
\citep{Alexiades96}.
To calculate the self-gravity, we adopted the Barnes-Hut tree algorithm
with opening angle of $\theta=0.5$ \citep{1986Natur.324..446B}. 
We do not use the individual time-steps and the particles are updated
simultaneously.

We adopted the tabulated EOS used in
\citet{2013ApJ...763....6T}, in which the internal degrees of freedom 
and chemical reactions of seven 
species ${\rm H_2,~H,~H^+,~He,~He^+,He^{++}, e^-}$ are included.
We assumed that the hydrogen and helium mass fractions 
were $X=0.7$ and $Y=0.28$, respectively.
The dust opacity table was obtained from \citet{2003A&A...410..611S} 
and the gas opacity table was obtained from \citet{2005ApJ...623..585F}.
The resistive model is described in \S 2.

We modelled the initial cloud core using an isothermal uniform gas sphere.
The initial mass and temperature of the core
were fixed at 1 $M_\odot$ and 10 K, respectively, with an
initial core radius of  $R \sim 3.0\times 10^3 $ AU. The core is
initially rigidly rotating
with an angular 
velocity of $\Omega_0=2.2 \times 10^{-13}~{\rm s^{-1}}$.
The product of the angular velocity and the free-fall time 
is $t_{\rm ff}\Omega_0=0.19$.
The initial magnetic field was uniform and parallel to the rotation ($z$-) axis 
with a 
strength of  $B_0=1.7\times 10^2 {\rm \mu G}$. 
The corresponding initial mass-to-flux ratio 
relative to the critical value was $\mu=(M/\Phi)/(M/\Phi)_{\rm crit}=4$ where
$\Phi=\pi R^2 B_0$. We adopted a  critical mass-to-flux 
ratio of  $(M/\Phi)_{\rm crit}=(0.53/3 \pi)(5/G)^{1/2}$ 
suggested by \citet{1976ApJ...210..326M}.
The initial cores were modelled with about $10^7$ SPH particles.
The boundary conditions of radiation transfer were introduced by fixing the
gas temperature to be 10 K when
the gas density was less than $2.0 \times 10^{-17} \cm$. 

We performed three simulations with and without Ohmic and ambipolar diffusion.
The model names and the diffusion processes included in each model are summarized in
Table 1.

\begin{table*}
\label{initial_conditions}

\begin{center}
\caption{The model names and the magnetic diffusion they include.
``Yes" means that the corresponding magnetic diffusion is included while a ``No" means that it is not. 
}		
\begin{tabular}{ccc}
\hline\hline
 Model  & Ohmic diffusion & Ambipolar diffusion \\
\hline
 1  & No & No \\
 2  & Yes& No \\
 3  & Yes& Yes \\
\hline
\end{tabular}
\end{center}
\footnotesize
\end{table*}

\section{Simulation results}
\subsection{Evolution at the centre of the cloud core}
\begin{figure}
\includegraphics[width=60mm,angle=-90]{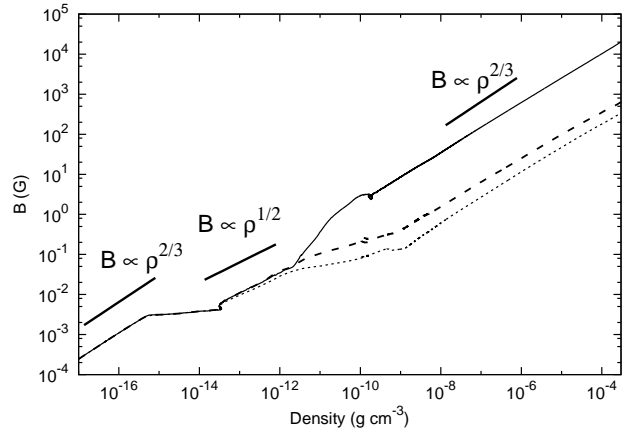}
\caption{
The evolution of the central magnetic field as a function of central density.
The solid, dashed, and dotted lines show the results of model 1 (ideal), 
model 2 (with Ohmic diffusion), and model 3 (with Ohmic and ambipolar diffusion), respectively.
Differences between the models can be seen when the central density 
exceeds $\rho_c>10^{-12} \cm$ and magnetic diffusion becomes effective.
}
\label{rho_B}
\end{figure}

To investigate how the magnetic field evolves during the gravitational collapse, 
we show the evolution of 
the central magnetic field as a function of the central density
in figure \ref{rho_B}.
At first, the magnetic field evolves as $B_c\propto \rho_c^{2/3}$.
This evolution is expected from a spherically symmetric collapse 
during which the central magnetic field evolves as $B_c \propto R^{-2}$
due to the conservation of the magnetic flux, 
where $R$ is the radius of the cloud. 
On the other hand, the central density
evolves as $\rho_c \propto R^{-3}$ or, equivalently, $R \propto \rho_c^{-1/3} $.
Thus, $B_c\propto R^{-2} \propto \rho_c^{2/3}$.
The increase in the magnetic field almost 
stops ($B_c \propto \rho_c^{0}$)  
at $10^{-15} \lesssim \rho_c \lesssim 10^{-14} \cm$
because the Lorentz force becomes comparable to the gravitational force
and the gas moves almost parallel to the magnetic field.
The $z$-component of the magnetic field still dominates other components and
the gas moves almost vertically.
As a result, a sheet-like structure (or pseudo disc) forms.
When the central density reaches $\rho_c \sim 10^{-13 } \cm$, 
the central magnetic field evolves as $B_c\propto \rho_c^{1/2}$ 
which indicates that the collapse becomes sheet-like. 
In the gravitationally collapsing isothermal 
sheet (whose scale-height is  $H=c_s^2/(G \Sigma)=c_s/\sqrt{G \rho_c}$ ),
the central magnetic field and density 
evolves as $B_c \propto R^{-2}$ and 
$\rho_c \propto R^{-2} H^{-1} \propto R^{-4}$ and hence $B_c \propto \rho_c^{1/2}$.

Once the central density exceeds $\rho \sim 10^{-12} \cm$, 
the magnetic diffusions becomes effective and the magnetic freezing 
is no longer valid for resistive models.
The magnetic flux is removed from the central 
part and the difference between the ideal model and
resistive models can be seen.
The central magnetic field of model 1 (the ideal model)
 is about sixty times larger than that of model 3 
(with Ohmic and ambipolar diffusion) when the central density reaches $\rho_c \sim 10^{-9} \cm$.
Around the  $\rho_c \sim 10^{-9} \cm$, the magnetic 
diffusion becomes ineffective again owing to the thermal ionization
and the flux freezing recovers in the resistive models.
This causes $B_c\propto \rho_c^{2/3}$ again.

\begin{figure*}
\includegraphics[width=45mm,trim=0mm 17mm 10mm 0mm,clip]{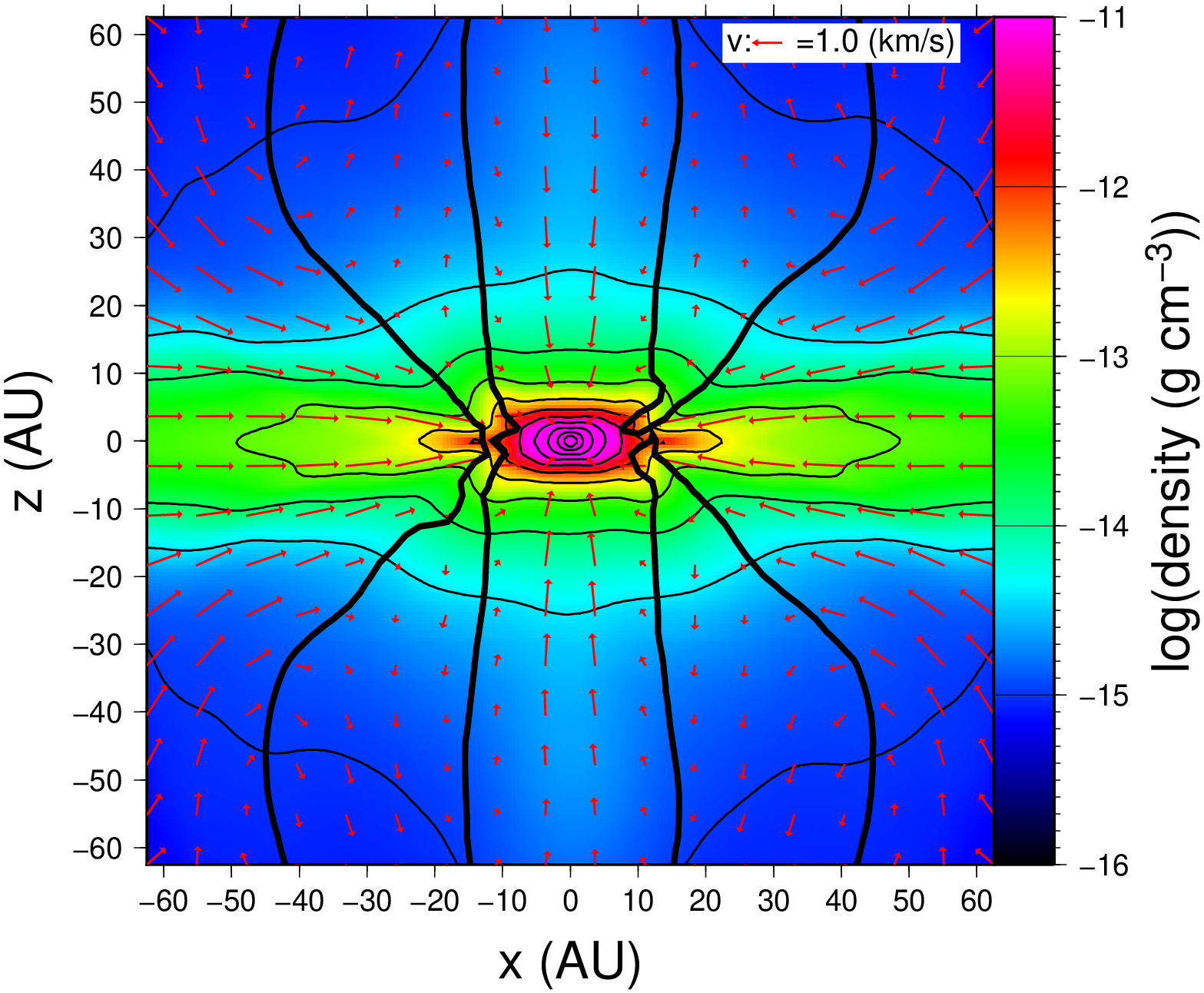}
\includegraphics[width=40mm,trim=15mm 17mm 10mm 0mm,clip]{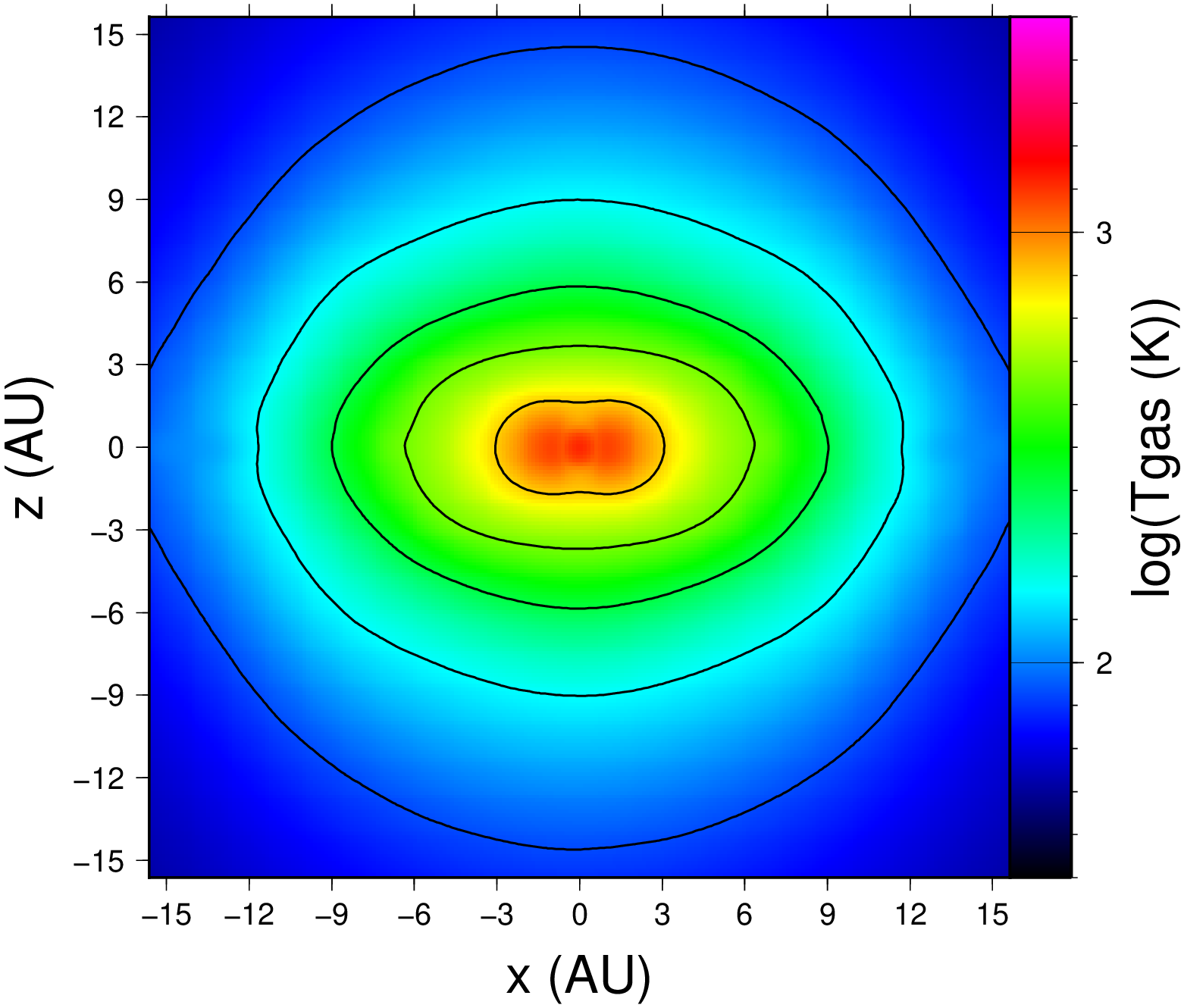}
\includegraphics[width=39.5mm,trim=20mm 17mm 10mm 0mm,clip]{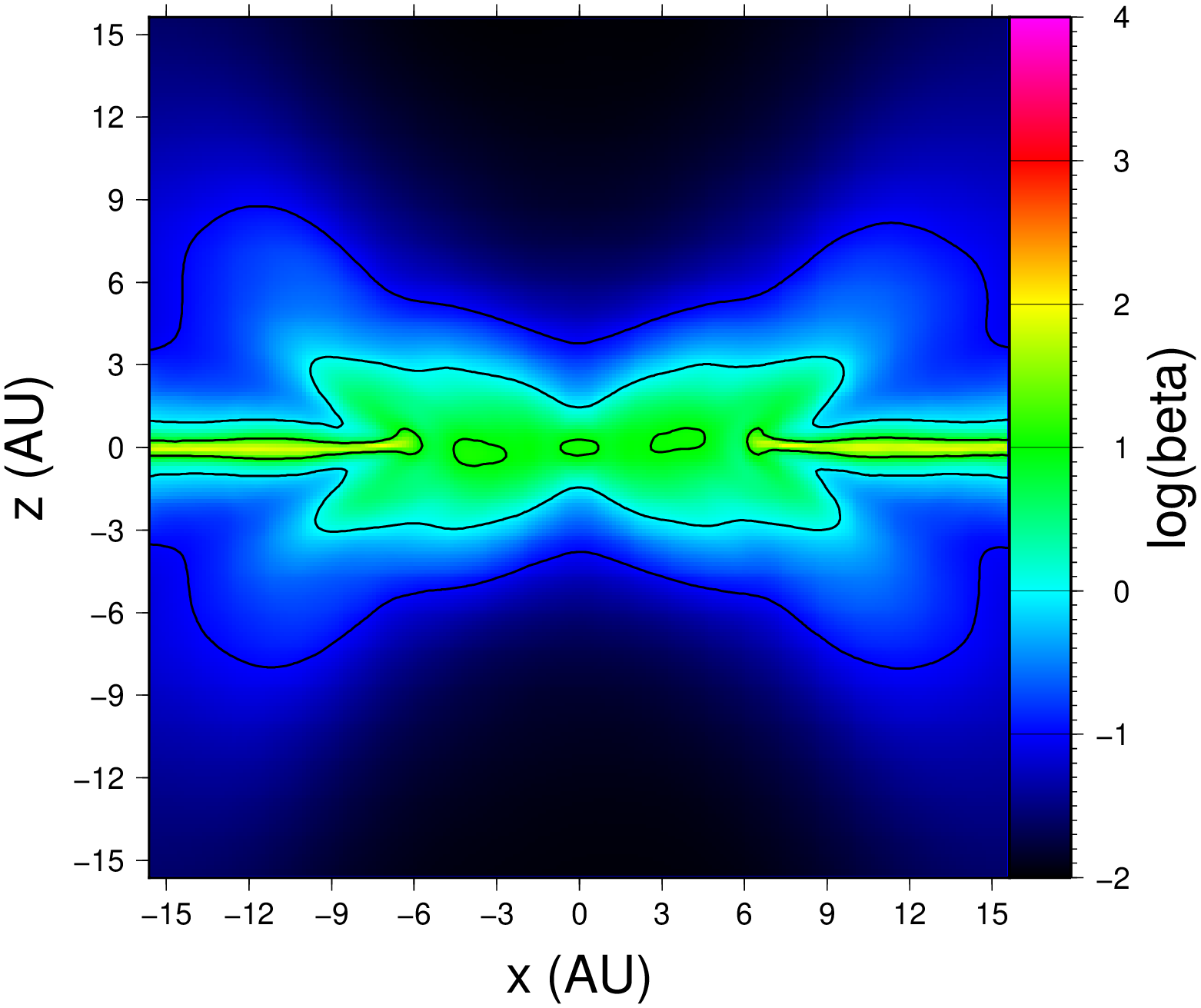}

\includegraphics[width=45mm,trim=0mm 17mm 10mm 0mm,clip]{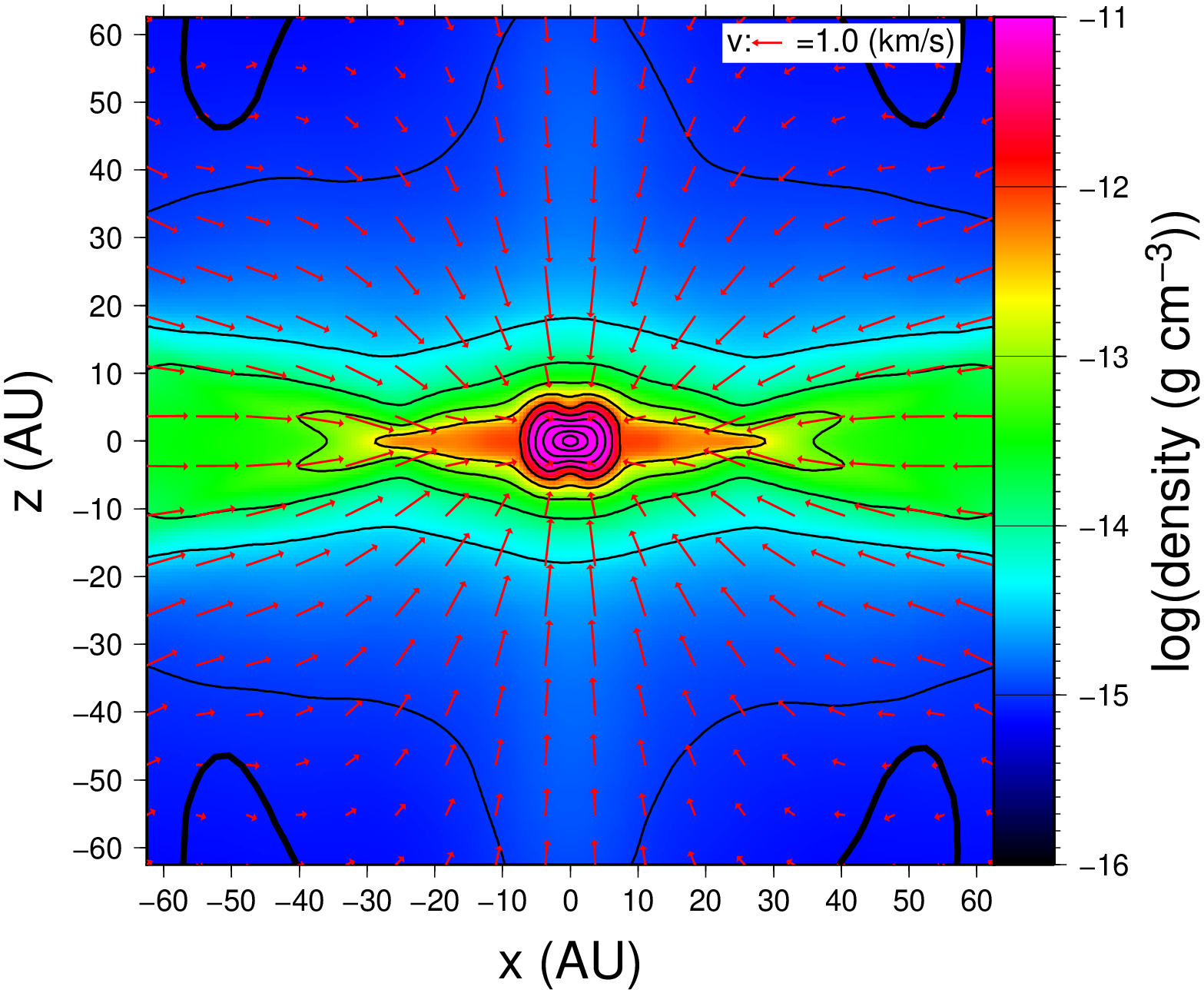}
\includegraphics[width=40mm,trim=15mm 17mm 10mm 0mm,clip]{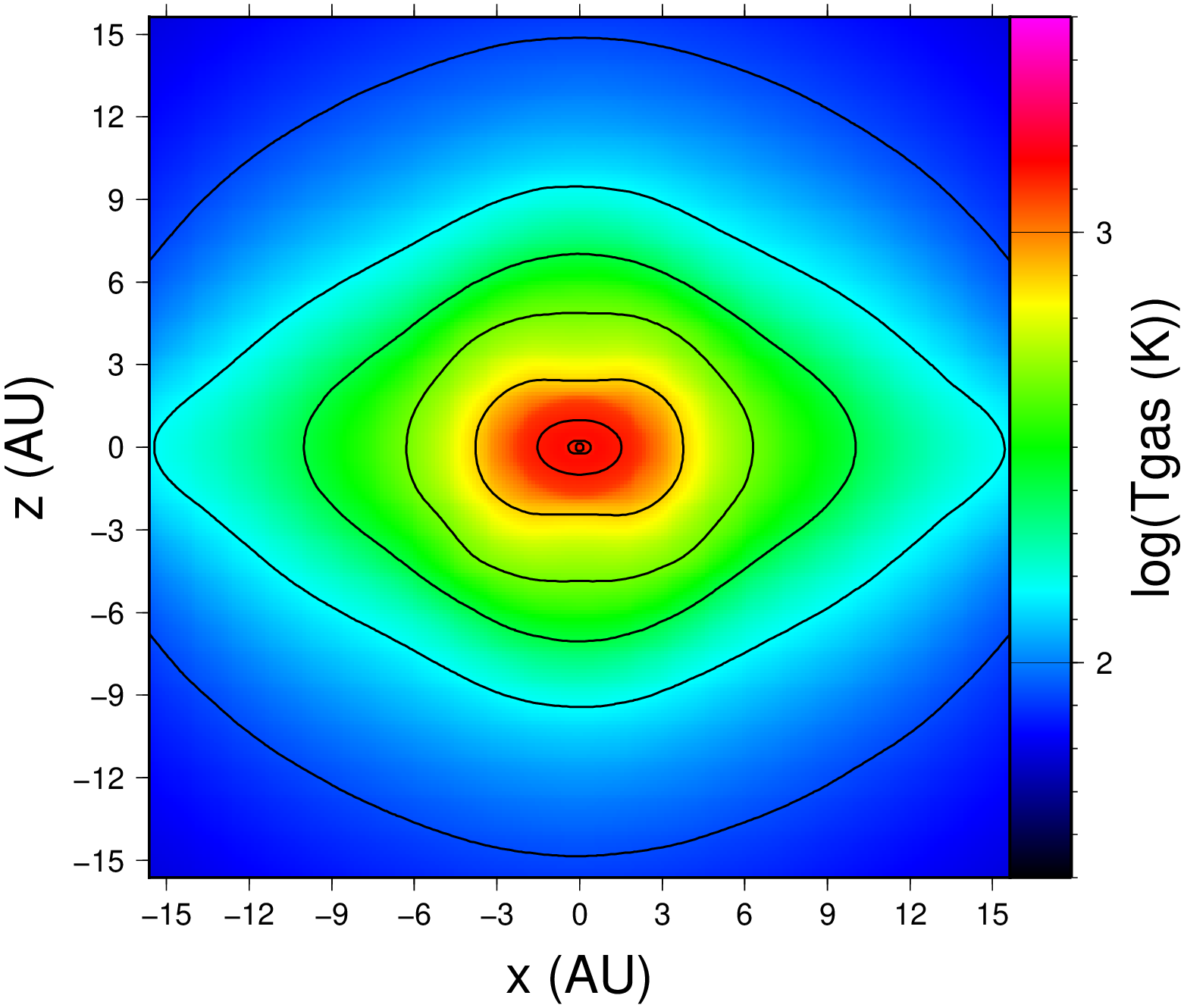}
\includegraphics[width=39.5mm,trim=20mm 17mm 10mm 0mm,clip]{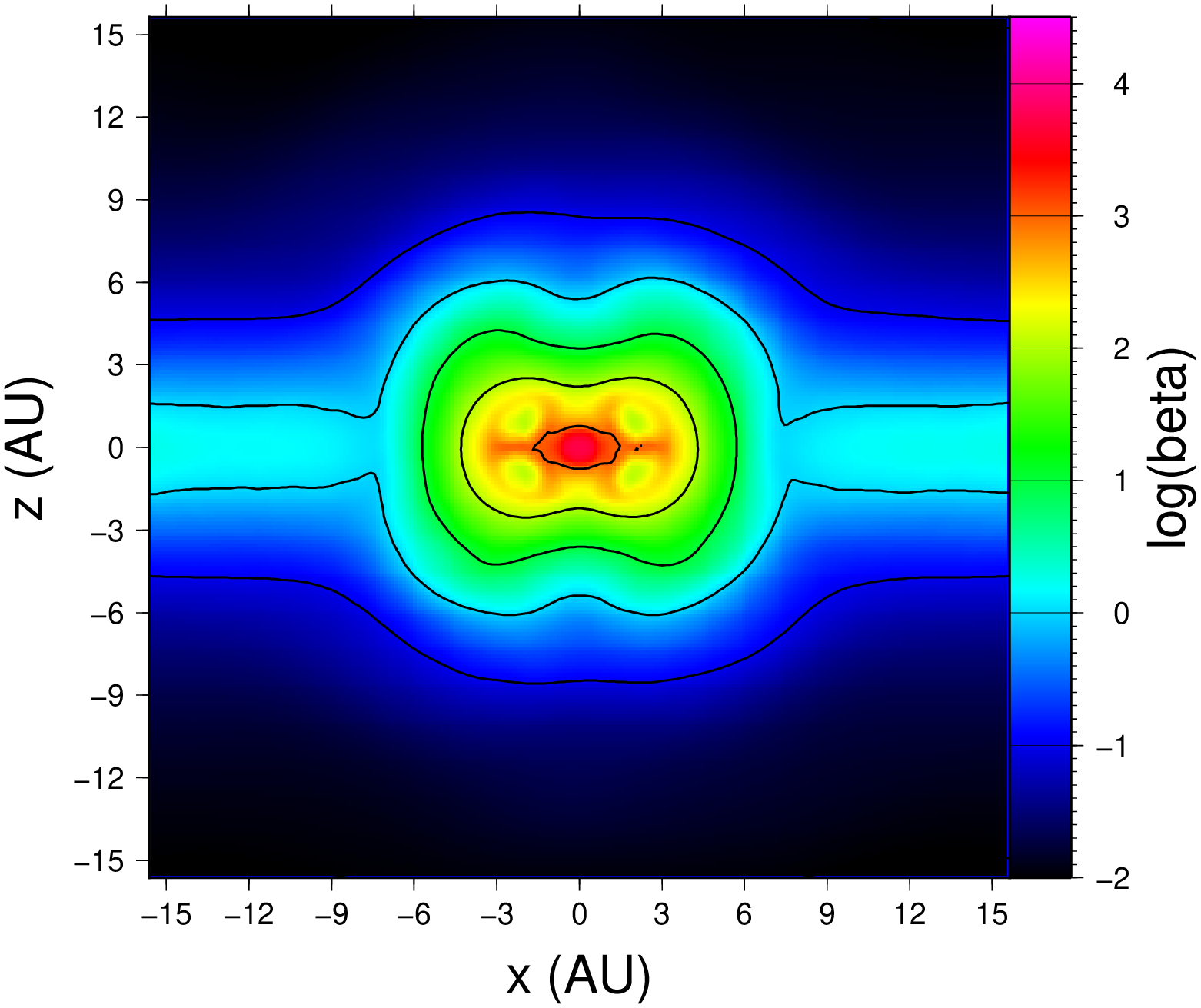}

\includegraphics[width=45mm,trim=0mm 0mm 10mm 0mm,clip]{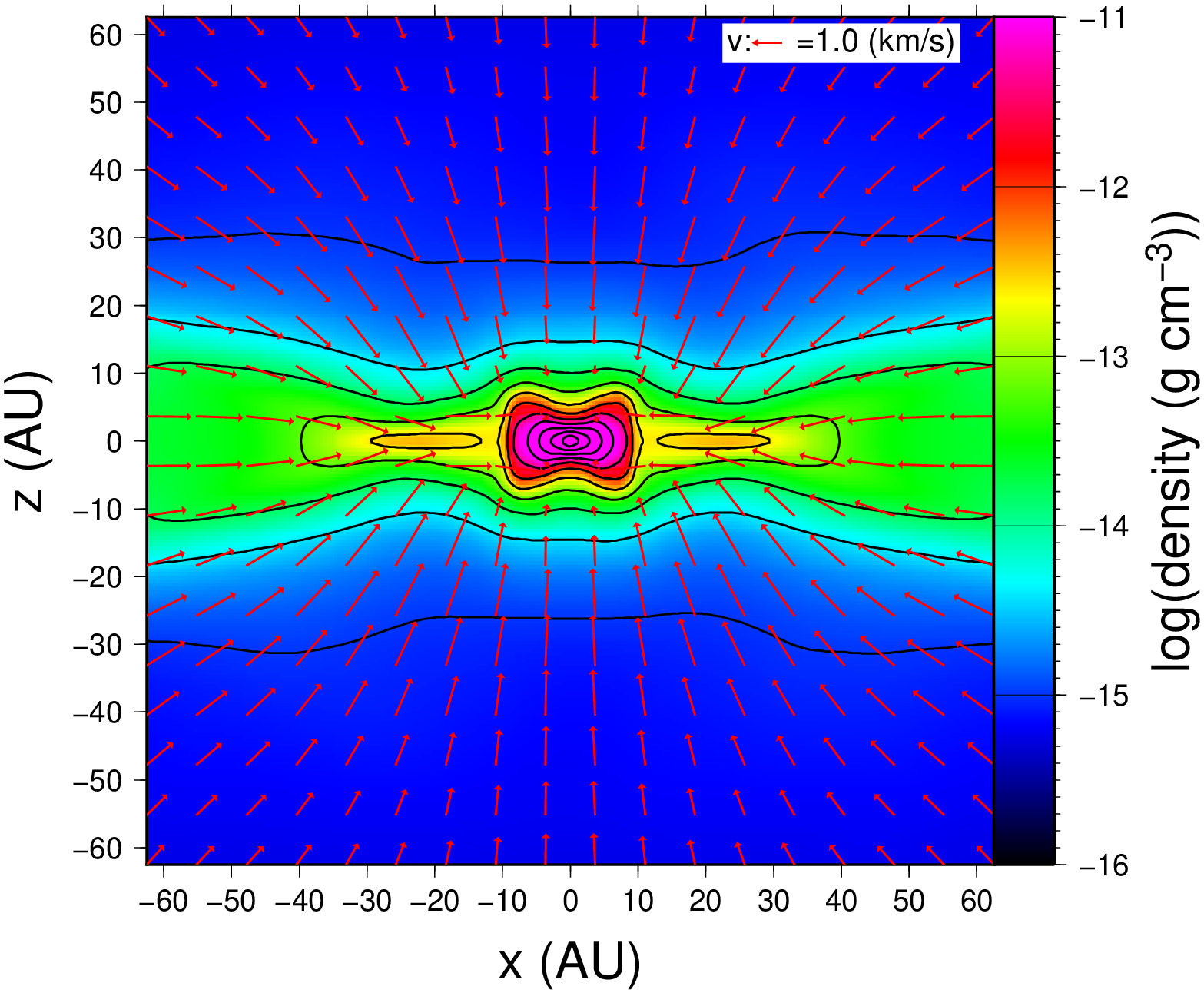}
\includegraphics[width=40mm,trim=15mm 0mm 10mm 0mm,clip]{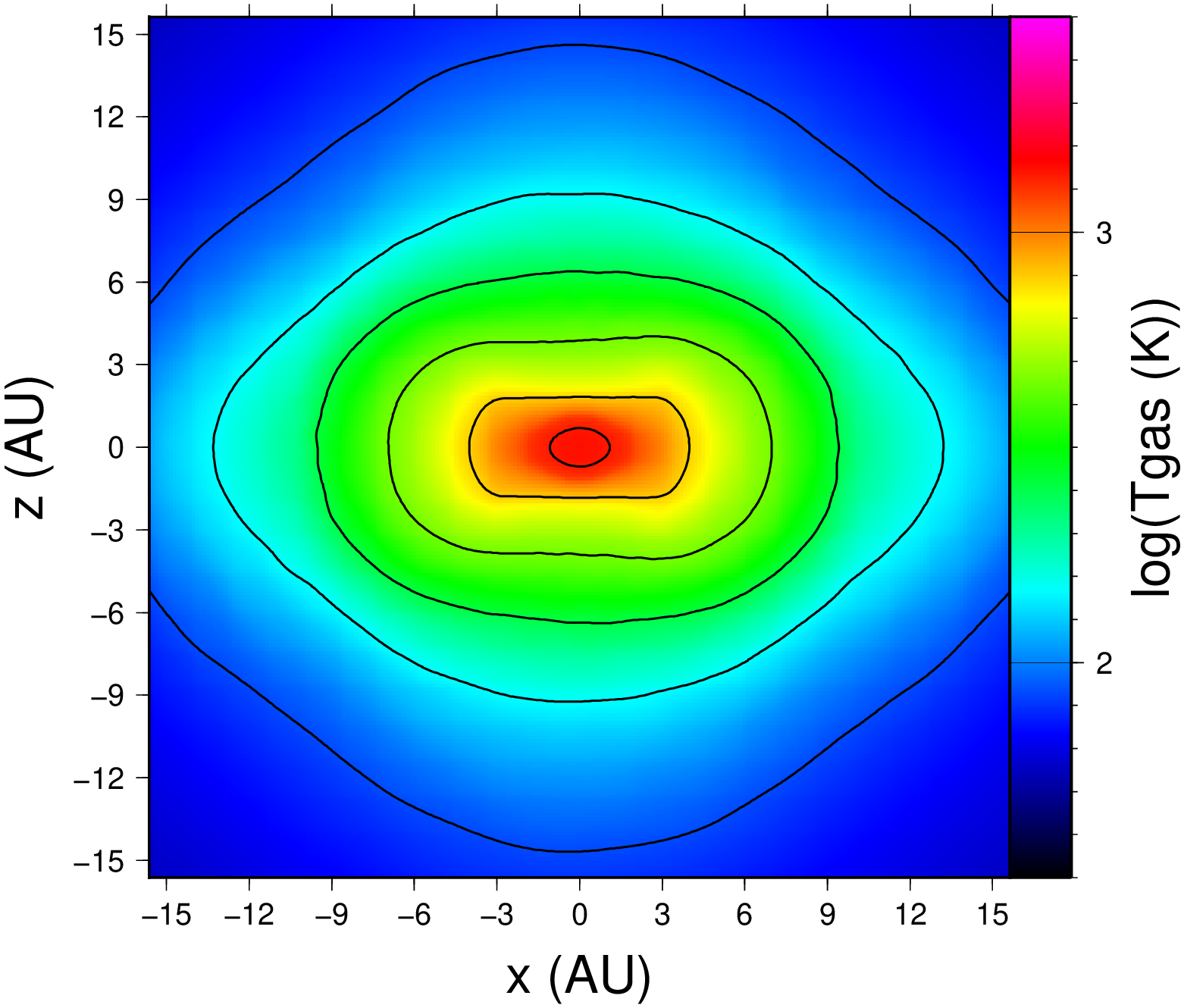}
\includegraphics[width=39.5mm,trim=20mm 0mm 10mm 0mm,clip]{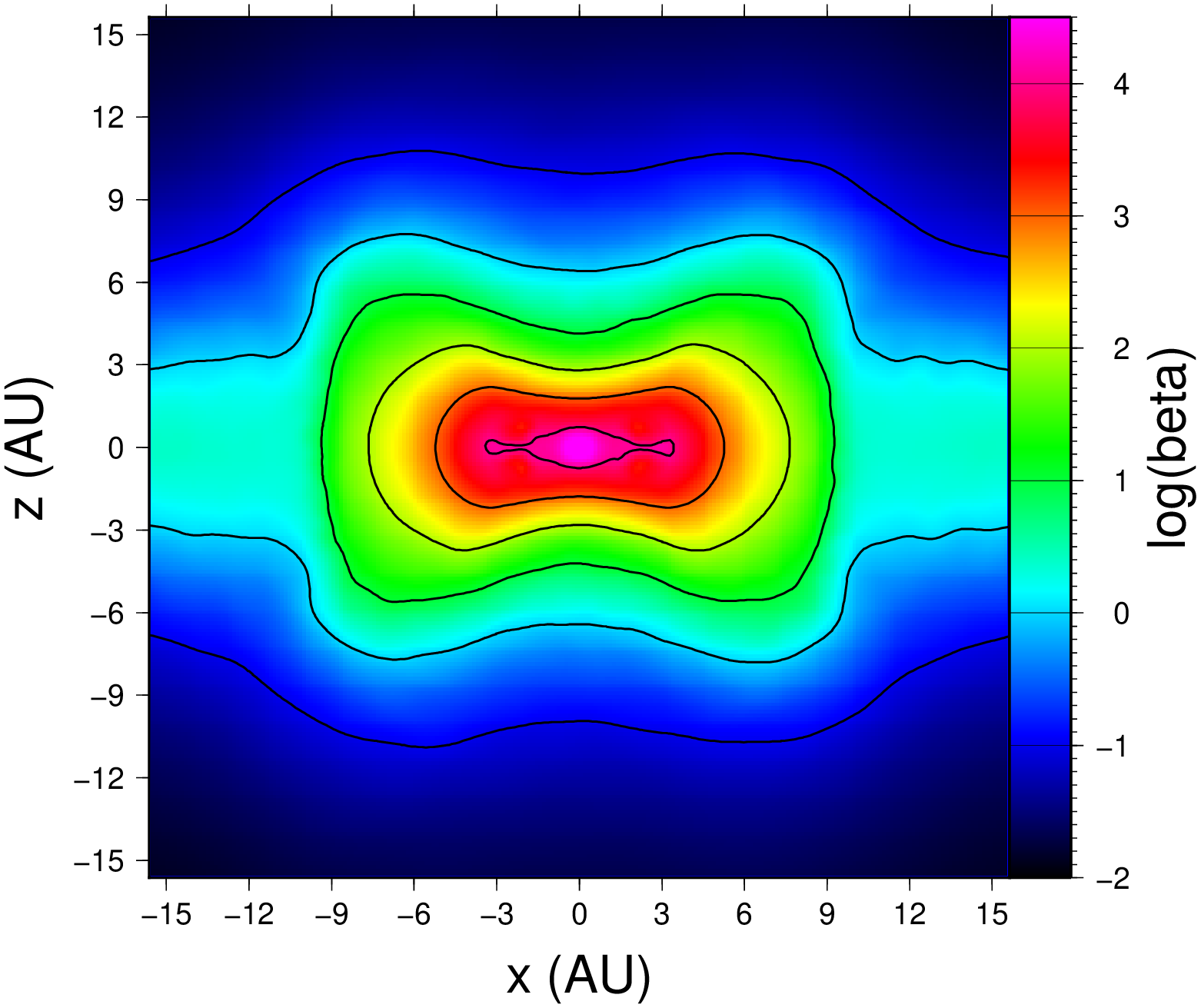}
\caption{
The cross sections of the density, 
gas temperature, and plasma $\beta$ (from left to right) 
around the first core in the $y=0$ plane.
The top row corresponds to model 1, the middle row to model 2, and the bottom row to model 3.
The thin black lines show the contour of each quantity, 
while the thick black lines in 
the density cross sections show the $|v_z|=0$ contour.
This traces the outflow regions.
The red arrows in the density cross sections show the velocity field.
The box size of the density cross sections is four times larger than the other
cross sections to show the outflow structures.
The color bars of the density, temperature, and plasma $\beta$ 
show $\log(\rho ~[\cm])$, $\log(T ~[{\rm K}]) $, and $\log(\beta)$, respectively.
}
\label{2dmap_around_firstcore}
\end{figure*}

\subsection{Structure of the first core}
When the central density reaches $\rho_c \sim 10^{-13} \cm$, 
the gas becomes opaque and the compressional heating 
due to the gravitational contraction cannot 
radiate away efficiently. As a result, the gas evolves adiabatically and
a pressure-supported core, the first core, forms. 
The first core phase lasts until the central temperature becomes
$T_c \sim 2000$ K (or $\rho_c\sim 10^{-8} \cm$) 
at which point the dissociation of hydrogen molecules begins.
The durations of the first core phase are
about 620 years for model 1,
810 years for model 2, and
940 years for model 3. The first core phase
is defined as the phase in which the central density is 
$10^{-13} \cm<\rho_c<10^{-8} \cm$.
The difference in the duration is due to
the rotation of the first core.

To investigate the structure  in and around the first core, 
we show the cross sections of the 
density, gas temperature, and plasma $\beta$ around the first core
in the $y=0$ plane in figure \ref{2dmap_around_firstcore}
at the end of the first core phase ($\rho_c \sim 3 \times 10^{-9} \cm$).
The plasma $\beta$ is defined as $\beta=P_{\rm gas}/P_{\rm mag}$ 
where $P_{\rm gas}$ and $P_{\rm mag}$ are the gas pressure and 
magnetic pressure, respectively.
Note that the box size of the density cross sections is four times 
that of the other
cross sections to compare the outflow structures of each model.

To obtain the cross section and the profiles,
the physical quantities are needed at grid points.
In this paper, the physical quantities are calculated  at grid points through,
\begin{eqnarray}
f(\mathbf{x}_{\rm grid})=\frac{\sum_j m_j \frac{f (\mathbf{x}_j)}{\rho_j}W(\mathbf{x}_{\rm grid}-\mathbf{x}_j,h_j)}{\sum_j m_j  \frac{1}{\rho_j} W(\mathbf{x}_{\rm grid}-\mathbf{x}_j,h_j)}.
\end{eqnarray}
In the left panels, we show the density cross section. 
The thick black solid lines show 
the $v_z=0$ contour and trace the outflow structure.
The outflow formed in both model 1 and 2, but did not
in model 3 at the epochs.
Although the outflow did not form in model 3, 
we confirmed that the outflow does form in a simulation 
with both Ohmic and ambipolar diffusion when the initial rotation of
the cloud core is slightly larger than in the model 3. 
Therefore, we conclude that the magnetic resistivity delays 
the formation of the outflow rather than suppressing it.
In our results, both the magneto-centrifugal force 
and the magnetic pressure play a role in driving the outflow.

In the middle panels, we show the temperature cross section 
around the first core.
The  high temperature  ($T \sim 1000$ K) regions with 
radius of $r\sim 5$ AU 
are formed at the centre due to the radiative transfer.
The high temperature region is extended compared 
to the case in which the barotropic EOS is adopted.
Because the thermal ionization becomes effective 
at $T\sim 1000$ K, the coupling between the magnetic
field and the gas recovers in the relatively large 
part of the first core when the radiative transfer is
taken into account.
This recoupling causes the amplification of the magnetic 
field inside the first core due to the rotation.

In the right panels, we show the cross section of the plasma $\beta$.
Because of the magnetic  diffusion, the magnetic flux is efficiently 
removed from the first core in the resistive models. 
Thus, in the resistive models, $\beta \gtrsim 10^3$ at the centre 
of the first core while in the ideal model, $\beta \sim 10$.
After the removal of the magnetic flux, the coupling between the gas and the 
magnetic field recovers at the central region of the first core
owing to the thermal ionization and the magnetic field in 
the first core is reamplified by the rotation.
As a result, the plasma $\beta$ around the centre slightly decreases 
in the resistive models. 
This amplification is clearly seen
in the middle right panel.

\begin{figure*}
\includegraphics[width=40mm,angle=-90]{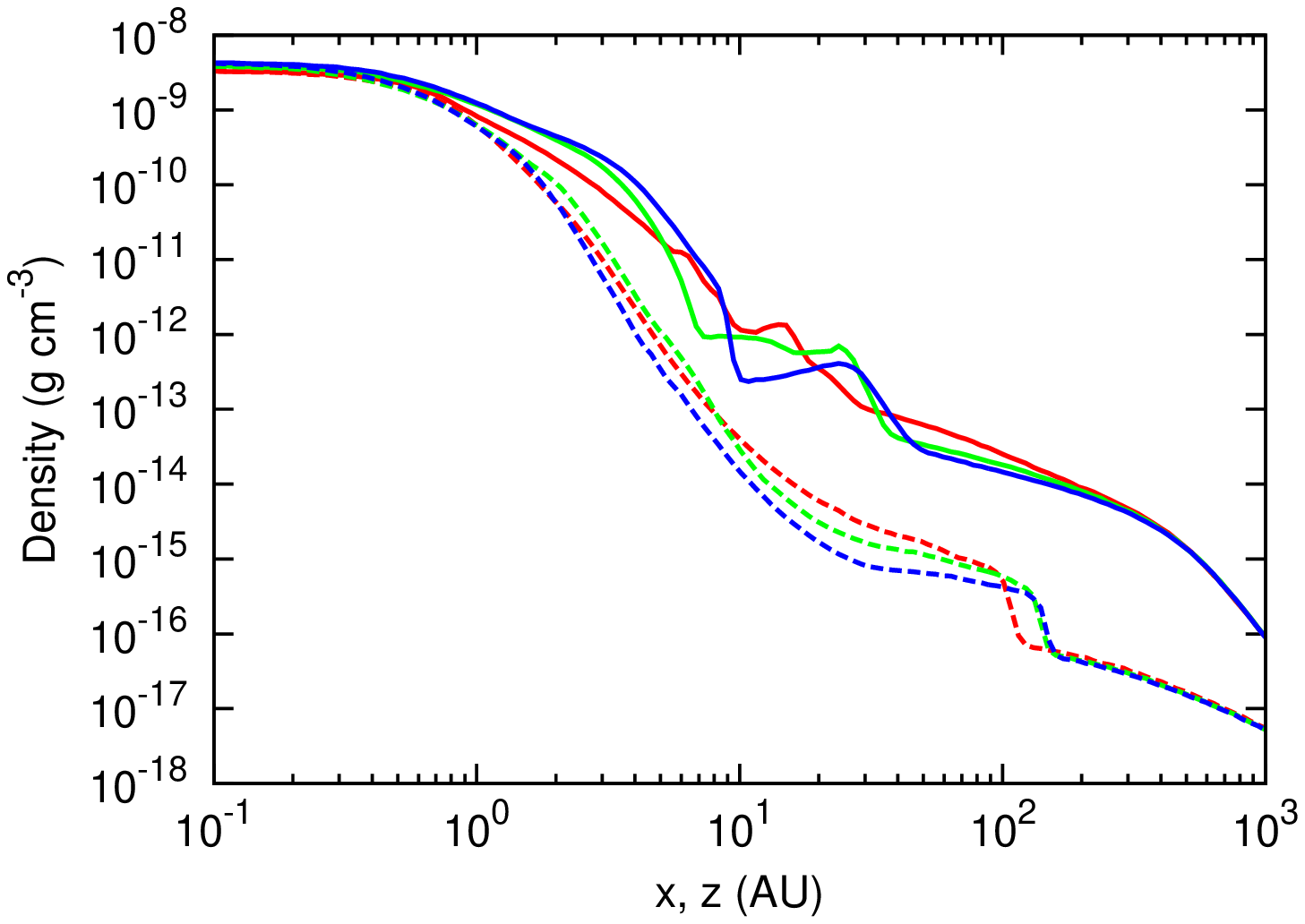}
\includegraphics[width=40mm,angle=-90]{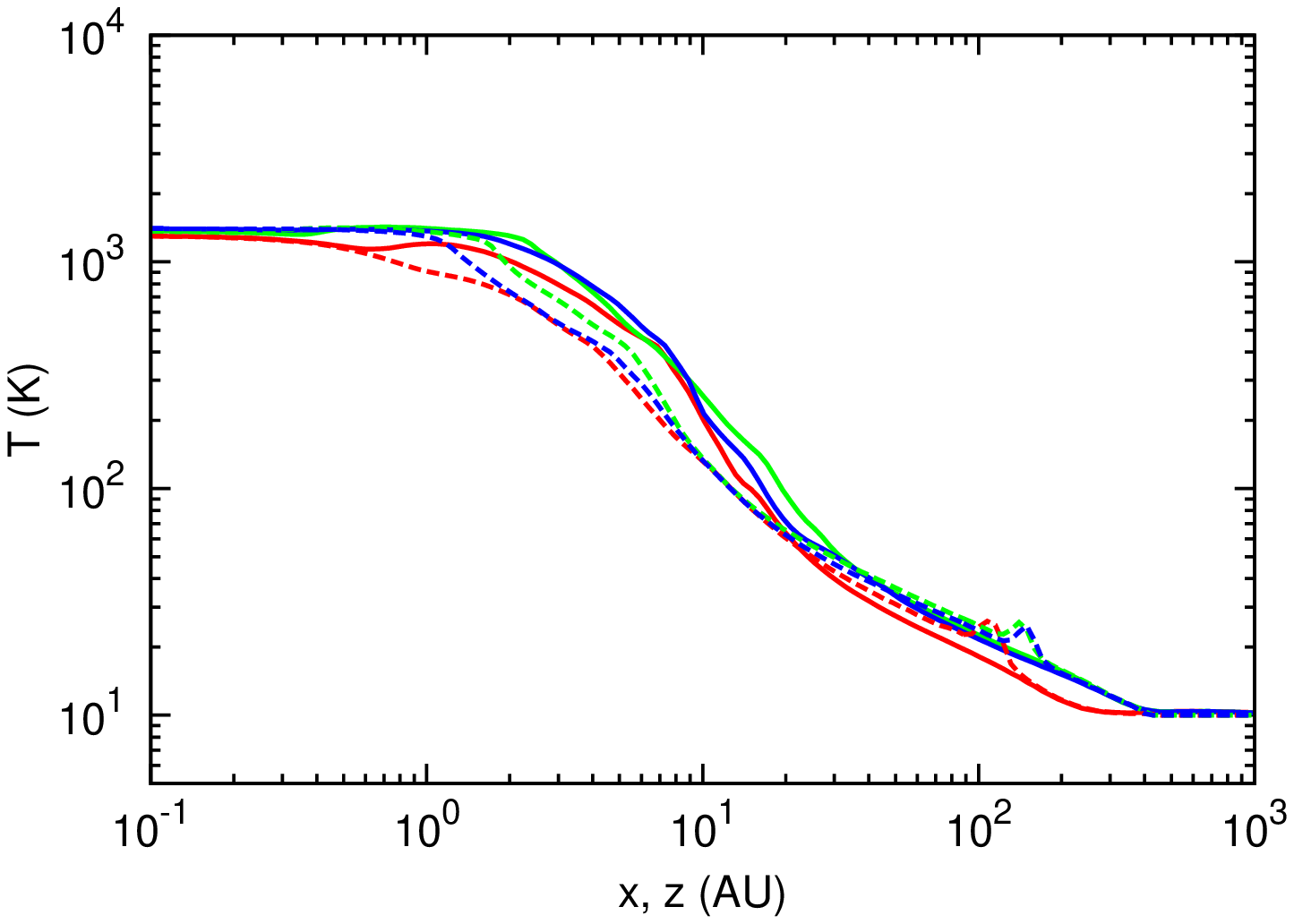}
\includegraphics[width=40mm,angle=-90]{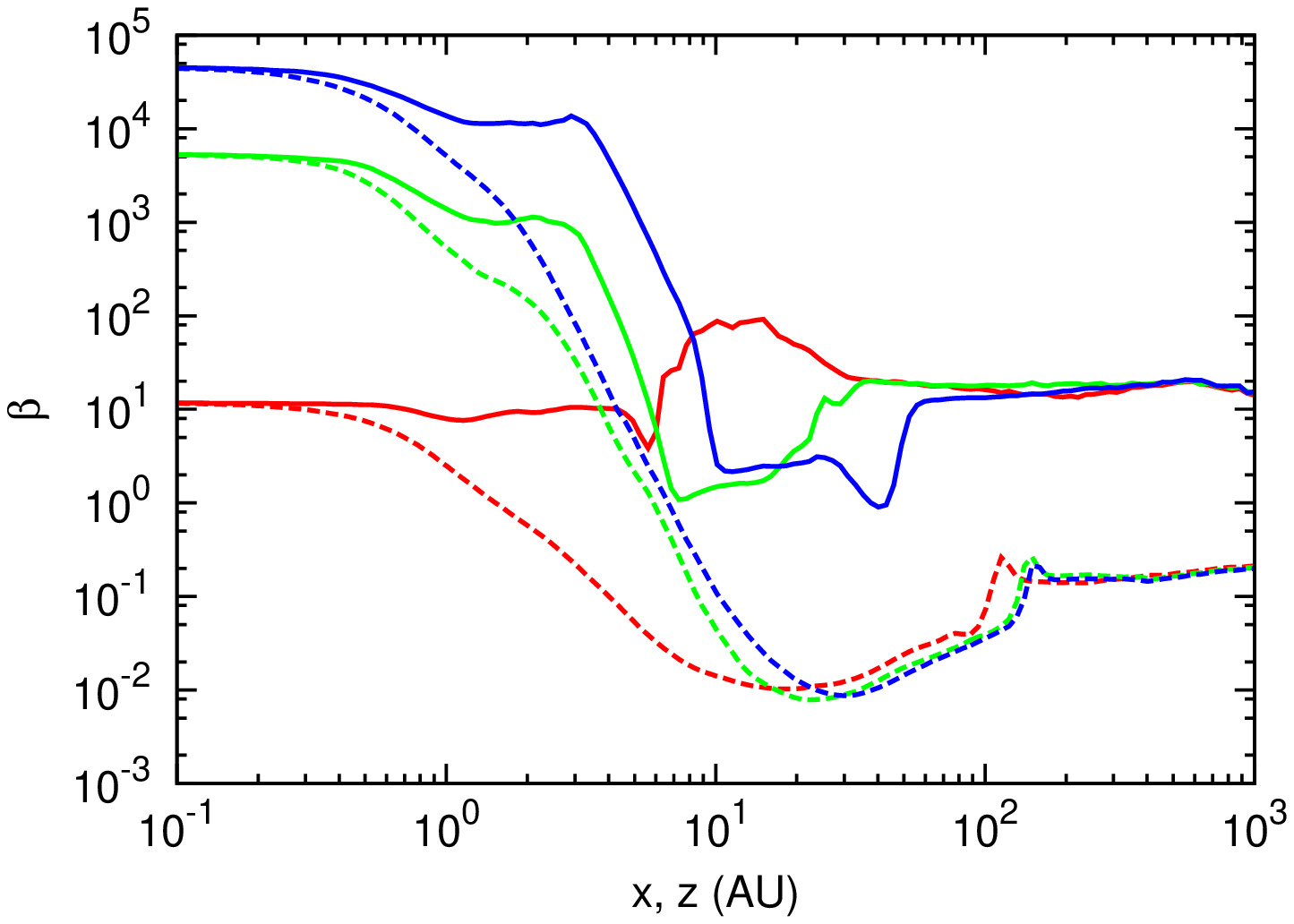}
\caption{
The density (left), gas temperature (middle), 
and plasma $\beta$ (right) profiles.
The epochs are the same as in the figure \ref{2dmap_around_firstcore}. 
The solid and dashed lines 
show the profiles of the x and z directions, respectively.
The red, green, and blue lines show 
the results of model 1 (ideal model),
model 2 (with Ohmic diffusion), 
and model 3 (with Ohmic and ambipolar diffusion), respectively.
}
\label{rho_and_Tgas_firstcore}
\end{figure*}

Figure \ref{rho_and_Tgas_firstcore} shows the 
profiles of the density, gas temperature, and plasma $\beta$ 
at the same epoch of figure \ref{2dmap_around_firstcore}.
In all models, the central density and the central temperature 
of the first core are $\rho_c \sim 3 \times 10^{-9} \cm$ 
and $T_c \sim 10^3 $ K, respectively. 
The density and temperature profiles show 
that the first cores formed in
each model have very similar structures.
This is because the angular momenta 
of the first cores are not significantly different and 
the structural difference caused by rotation is negligible. 
The density on the $x$-axis is larger than that on the $z$-axis 
outside of the first core
because the pseudo disc has formed in the x direction. 
On the other hand, the temperature profiles
along the $x$ and $z$-axis do not differ significantly 
and the temperature structure is hence almost spherically symmetric.

Due to the magnetic diffusions, the plasma $\beta$ in the central
region of the first core differs significantly between the ideal 
model and the resistive models. 
In model 1, the plasma $\beta$ inside the first core
is $\beta \sim 10$ and almost constant in the x direction. 
In the model 2,
the plasma $\beta$ at the centre of the first core becomes $\beta  \sim 6 \times 10^3$. 
This is hence about three orders magnitude greater than for the ideal model.
The magnetic flux removed from the first core piles 
up around it and the plasma $\beta$ on the $x$-axis  
becomes smaller than the ideal model at the perimeter of the first core 
($x\sim 10$ AU).
In model 3,
the plasma $\beta$ at the centre of the first core 
becomes $\beta \sim 6 \times 10^4$, which is much higher than for the model 2. 
In the z direction, the plasma $\beta$ quickly decreases in all models
because of the large density gradient in this direction
and the magnetic field amplification by the first core rotation.
{\rm
Because the plasma $\beta$ is larger than 10 inside the first core,
the magnetic pressure does not affect the pressure support in the first
core.
}

A notable difference between models 2 and 3 is 
the plasma $\beta$ in the x direction {\it at the perimeter of} the first core.
In the model 2, only Ohmic diffusion is considered. 
The Ohmic diffusion coefficient 
is an increasing function of density and does not 
depend on the magnetic field. 
Roughly speaking, the Ohmic diffusion does not 
play a role when $\rho \lesssim 10^{-13} \cm$ \citep[][]{2007ApJ...670.1198M}.
Because the density of the first core is $\rho \gtrsim 10^{-13} \cm$, 
the magnetic flux piles up outside the first core. By this pile-up, 
the plasma $\beta$ beyond the first core 
in model 2 is $\beta \sim 1$ around $x=10$ AU and becomes much smaller than
for model 1 at larger $x$.
In model 3, the ambipolar diffusion is included as well. 
The diffusion coefficient of the ambipolar diffusion
is a function of the magnetic field and does 
not depend strongly on the density. 
Therefore, it is expected that the pile-up of the magnetic flux 
is smoothed by the ambipolar diffusion. 
Actually, the region of small plasma 
$\beta$ ($\beta \sim 1$) in the x direction broadens in the right panel.
This difference can also be seen in the right panels of 
figure \ref{rho_and_Tgas_firstcore}.

\begin{figure*}
\includegraphics[width=40mm,angle=-90]{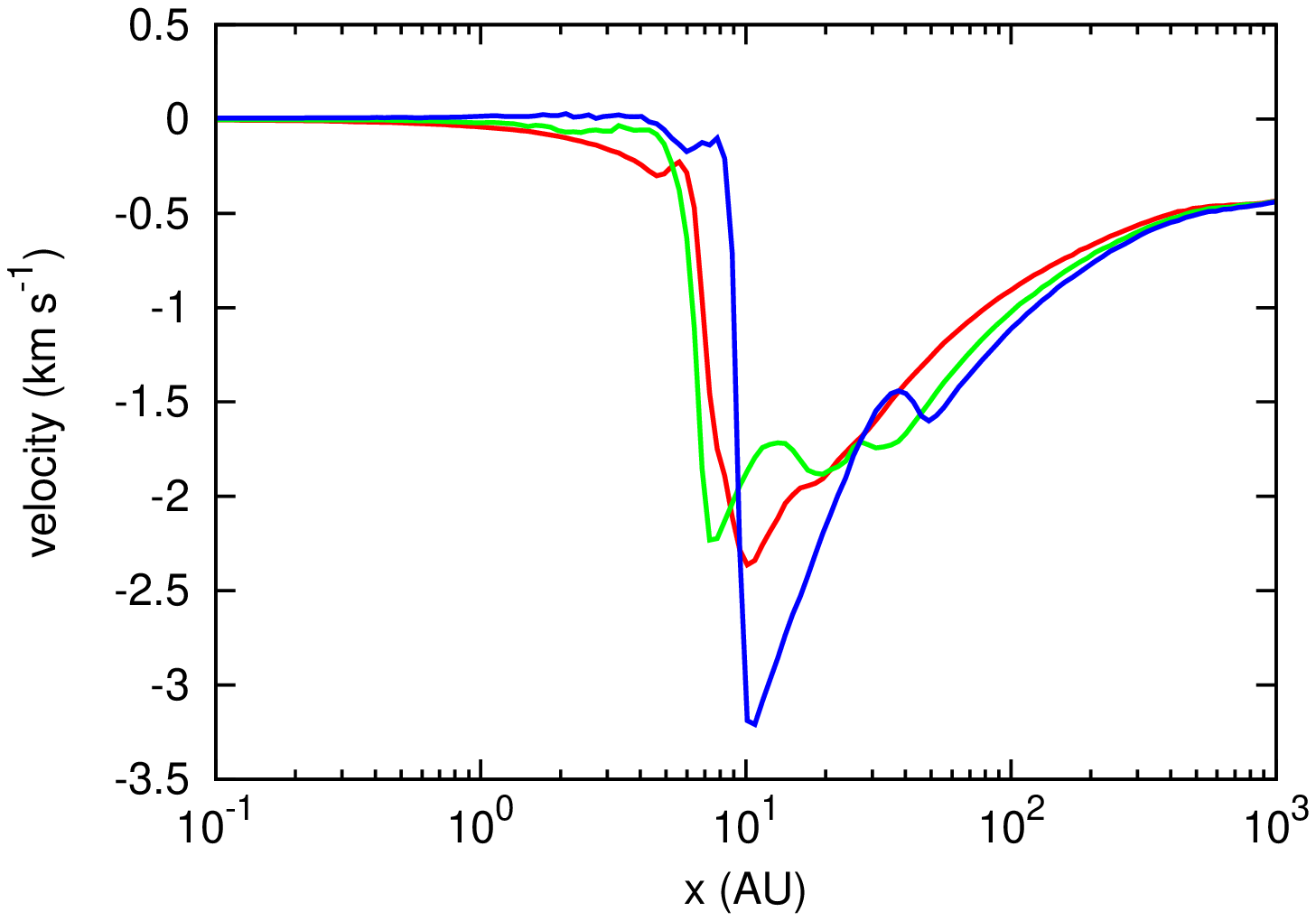}
\includegraphics[width=40mm,angle=-90]{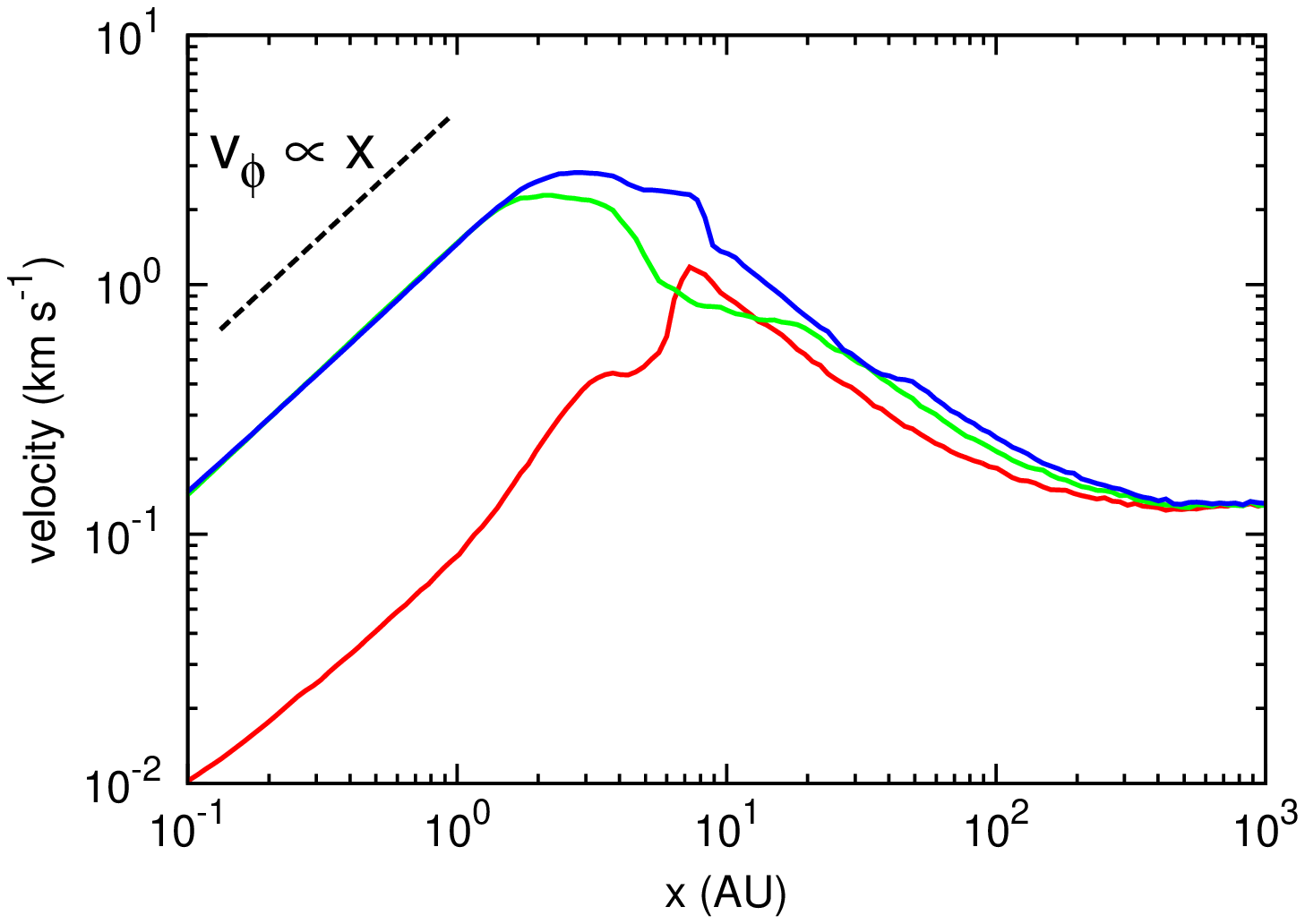}
\includegraphics[width=40mm,angle=-90]{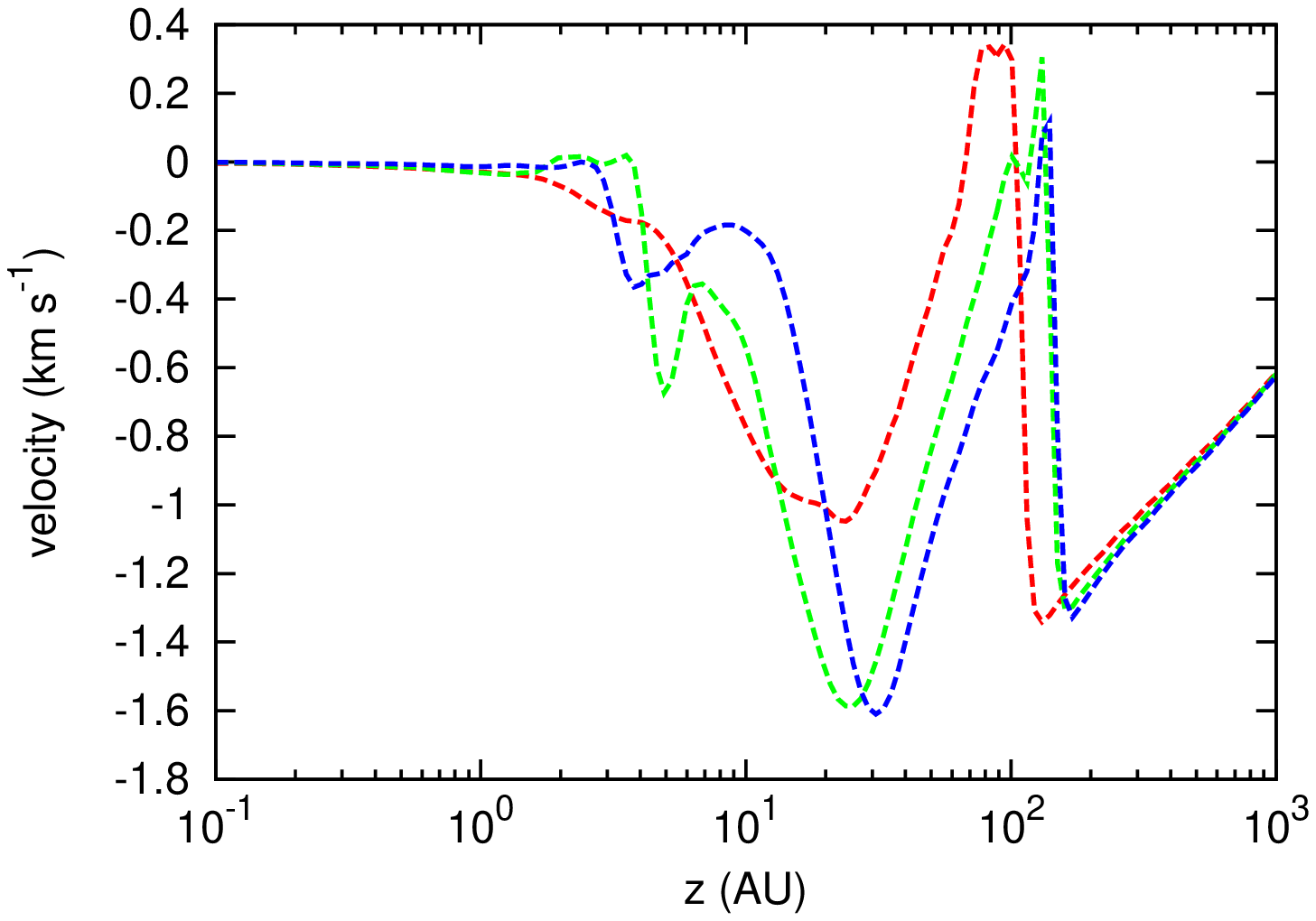}
\caption{
The profiles of the infall velocity (left) and rotation velocity (middle)
in the x direction and the infall velocity in the z direction (right).
The epochs are the same as in the figure \ref{2dmap_around_firstcore}.
The red, green, and blue lines show 
the results of model 1 (ideal model),
model 2 (with Ohmic diffusion), 
and model 3 (with Ohmic and ambipolar diffusion), respectively.
}
\label{vx_and_vy_firstcore}
\end{figure*}

In figure \ref{vx_and_vy_firstcore}, we show the infall and 
rotation velocities along the $x$-axis and the infall velocity along the $z$-axis.
The infall velocity along 
the $x$-axis is larger than that along the $z$-axis
and the density on the $x$-axis is also much higher 
than on the $z$-axis at the surface of the first core ($x,z \sim 10$ AU). 
Therefore, the mass accretion onto the first core is asymmetric 
and is maximal in the horizontal direction.

The rotation velocity $v_{\phi}$ reaches its 
maximum value at $x \sim 2$ AU in models 2 and 3.
Inside this radius, the velocity profile 
obeys the rigid rotation relation,
$v_{\phi} \propto x$.
Note that a rigid rotation is expected when the
density is constant because 
$v_{\phi}\propto \sqrt{M_r/r}\propto \sqrt{\rho_0 r^3/r}\propto r$, where
$M_r$ and $\rho_0$ are the mass inside $r$ and the density, respectively.
In model 2, $v_{\phi}$ sharply decreases at the $r\sim 4$ AU. 
This is caused by the strong magnetic braking by the piled-up magnetic
field.
As mentioned above, the magnetic flux piles up around 
the first core.
Hence, the magnetic braking is locally enhanced at $r\sim 4$ AU
and the rotation velocity is decreases.
The profile of the model 1 also obeys the relation 
of $v_{\phi} \propto x$ for $x\lesssim1$ AU. On the other hand,
for $3 ~{\rm AU}\lesssim x \lesssim 10$ AU, 
the profile has a complex structure. 
This structure is also caused by the magnetic  braking. 
Note that the plasma $\beta$ inside the first core 
is still small in model 1.

\begin{figure}
\includegraphics[width=60mm,angle=-90]{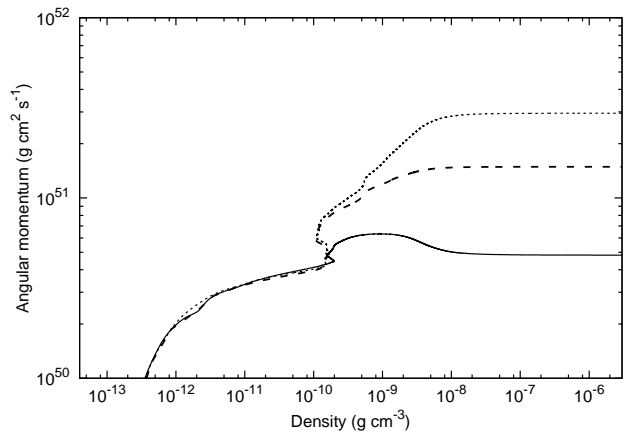}
\caption{
The evolution of the angular momentum of the first core in relation to
the central density.
The solid, dashed, and dotted lines show 
the results of model 1, 2, and 3, respectively.
}
\label{rho_J13}
\end{figure}

In figure \ref{rho_J13}, we show the evolution of the angular momentum 
of the first core in relation to the central density.
We define the first core as the region where $\rho>10^{-13} \cm$.
As we have seen above, the magnetic field in the first core 
becomes weak due to the magnetic diffusion which causes
an inefficient angular momentum transfer by the magnetic braking.
Thus, it is expected that the angular momentum of the first core becomes 
large in resistive models and it indeed 
becomes large when the magnetic diffusion is included.
The difference in the angular 
momentum between model 1 and model 3 
is a factor of 6 and hence, insignificant.
Most of the initial angular momentum of the fluid element has 
already been removed during the isothermal collapse phase.
With the parameters adopted in our simulations,
a disc of $r\sim 100$ AU forms 
when the magnetic field is neglected
\citep[see, e.g.,][]{2011MNRAS.416..591T,2015MNRAS.446.1175T}.
Therefore, we conclude that the angular momentum of the first core 
depend more strongly on the initial condition of the molecular cloud cores 
\citep[see, e.g.,][]{2012A&A...543A.128J}.

\subsection{Formation of the protostar}
\begin{figure*}

\includegraphics[width=47mm,trim=0mm 10mm 10mm 0mm,clip]{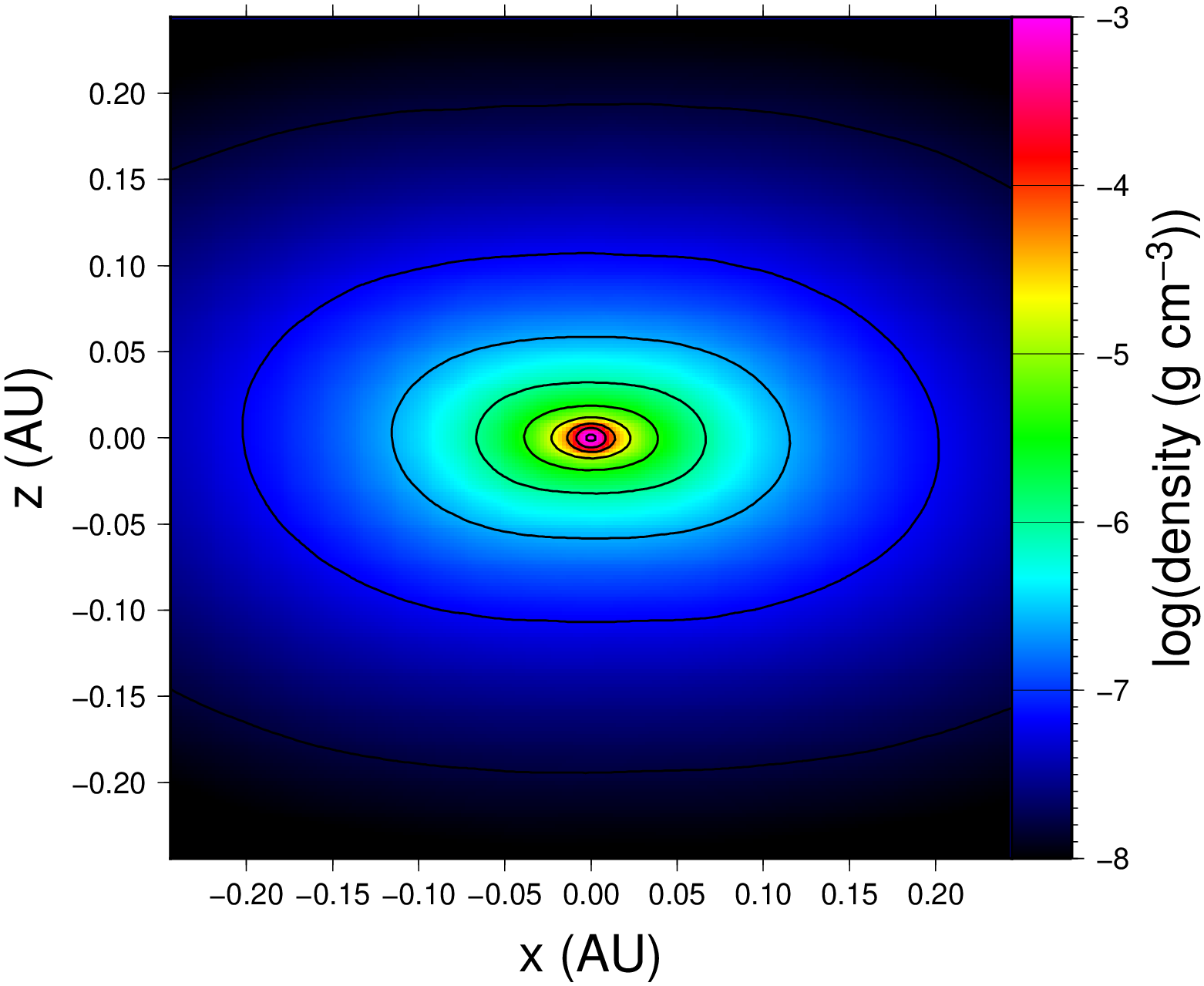}
\includegraphics[width=40mm,trim=25mm 10mm 10mm 0mm,clip]{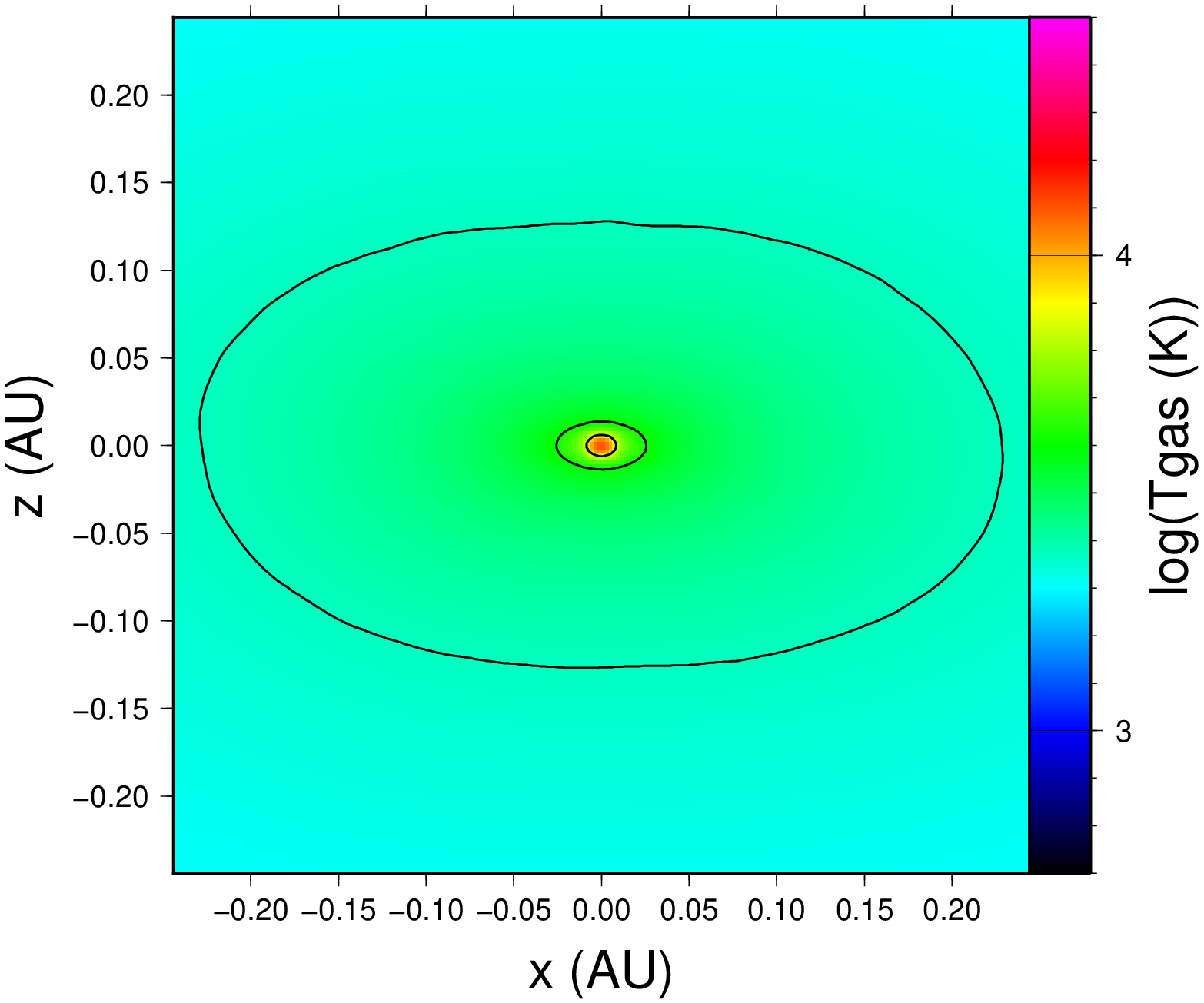}
\includegraphics[width=40.5mm,trim=25mm 10mm 10mm 0mm,clip]{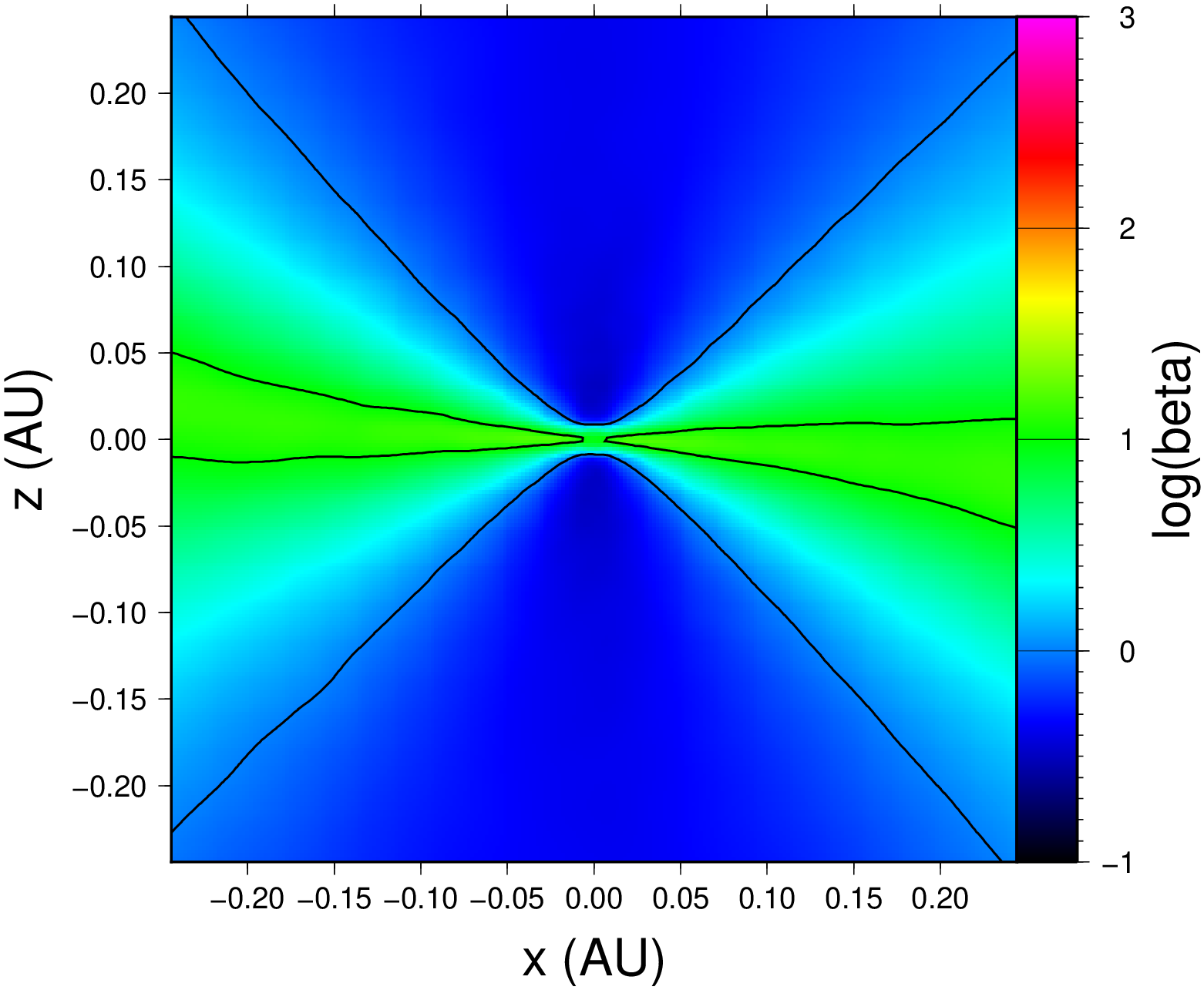}

\includegraphics[width=47mm,trim=0mm 17mm 10mm 0mm,clip]{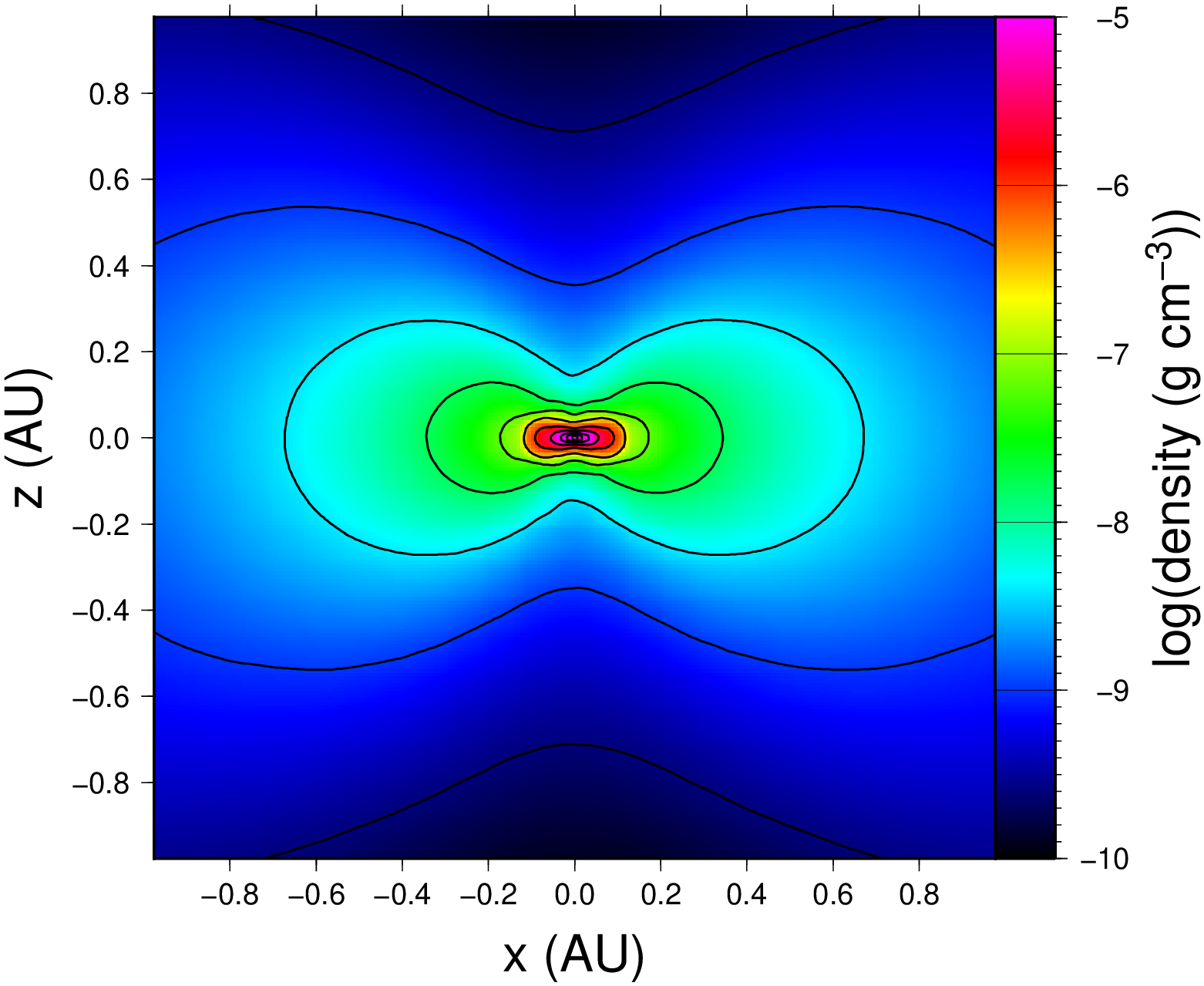}
\includegraphics[width=40mm,trim=23mm 17mm 10mm 0mm,clip]{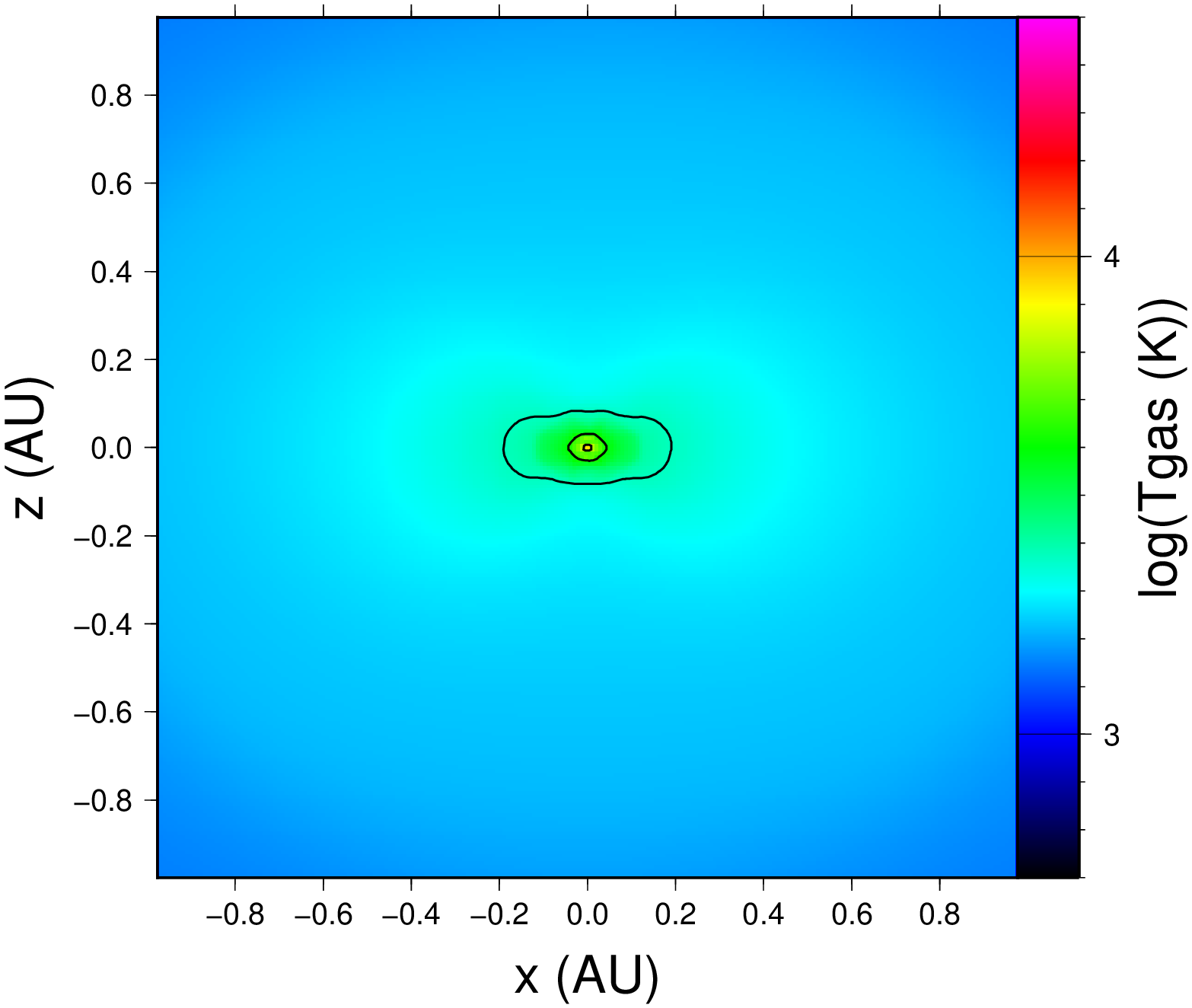}
\includegraphics[width=40mm,trim=23mm 17mm 10mm 0mm,clip]{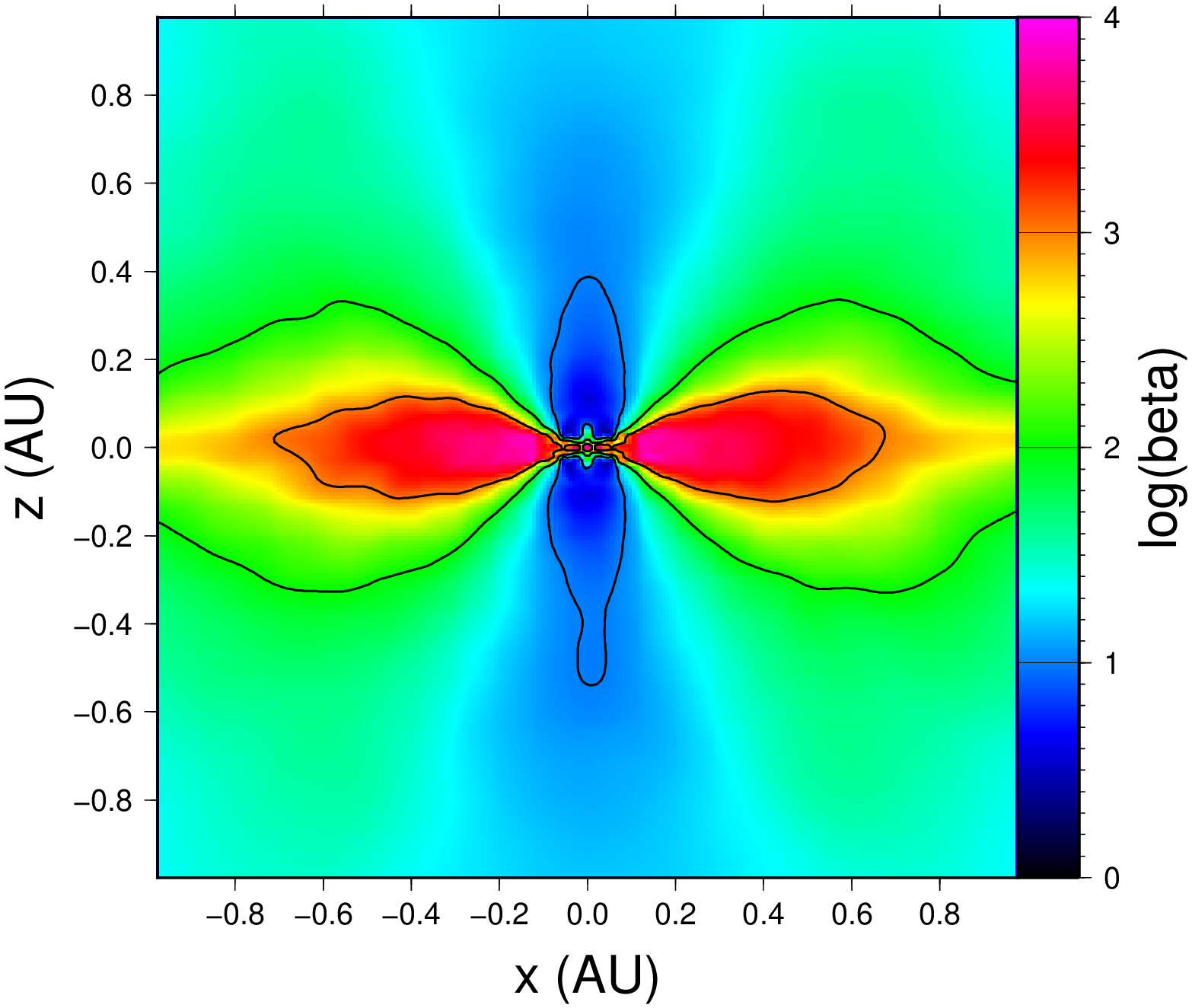}

\includegraphics[width=47mm,trim=0mm 0mm 10mm 0mm,clip]{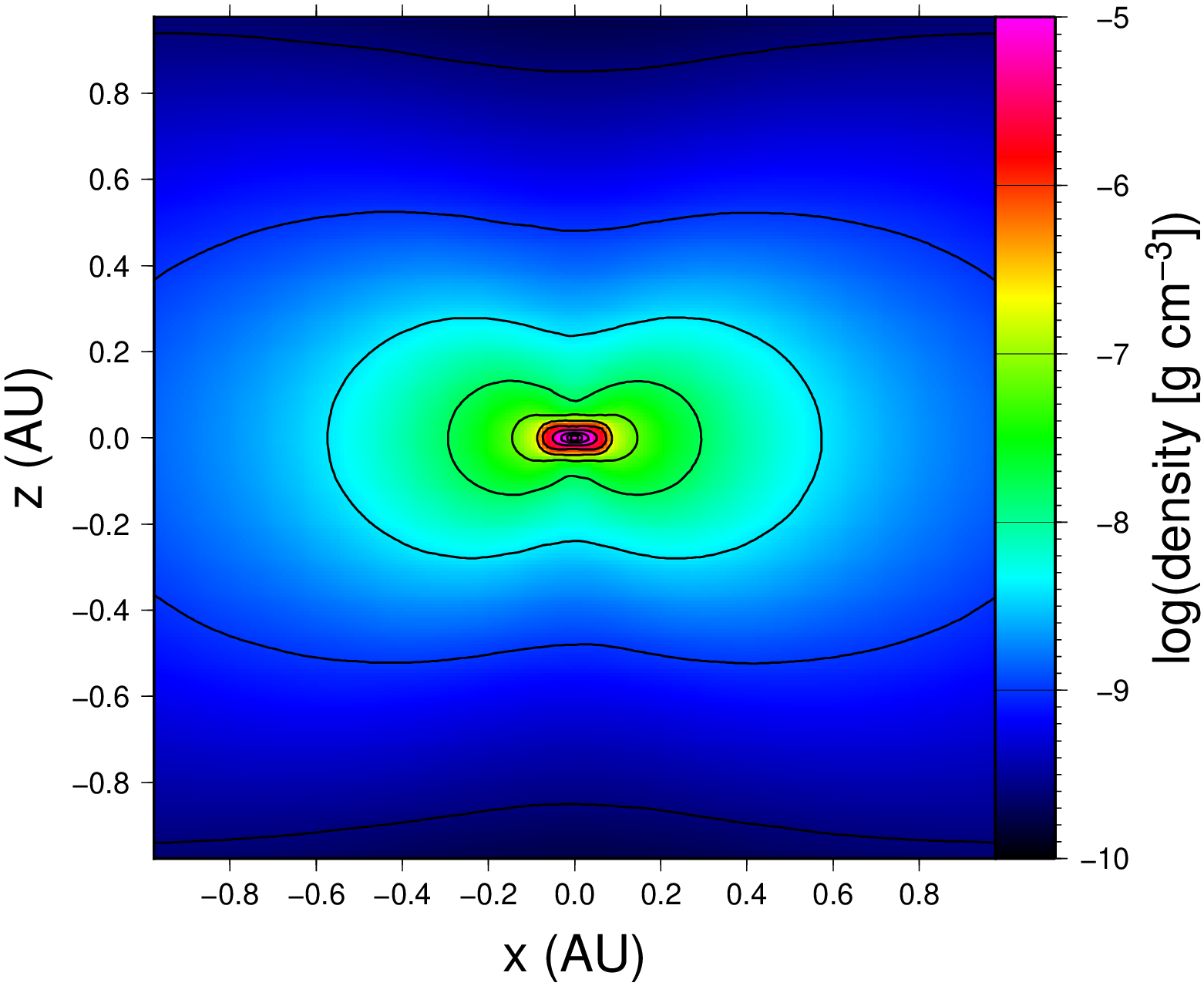}
\includegraphics[width=40mm,trim=23mm 0mm 10mm 0mm,clip]{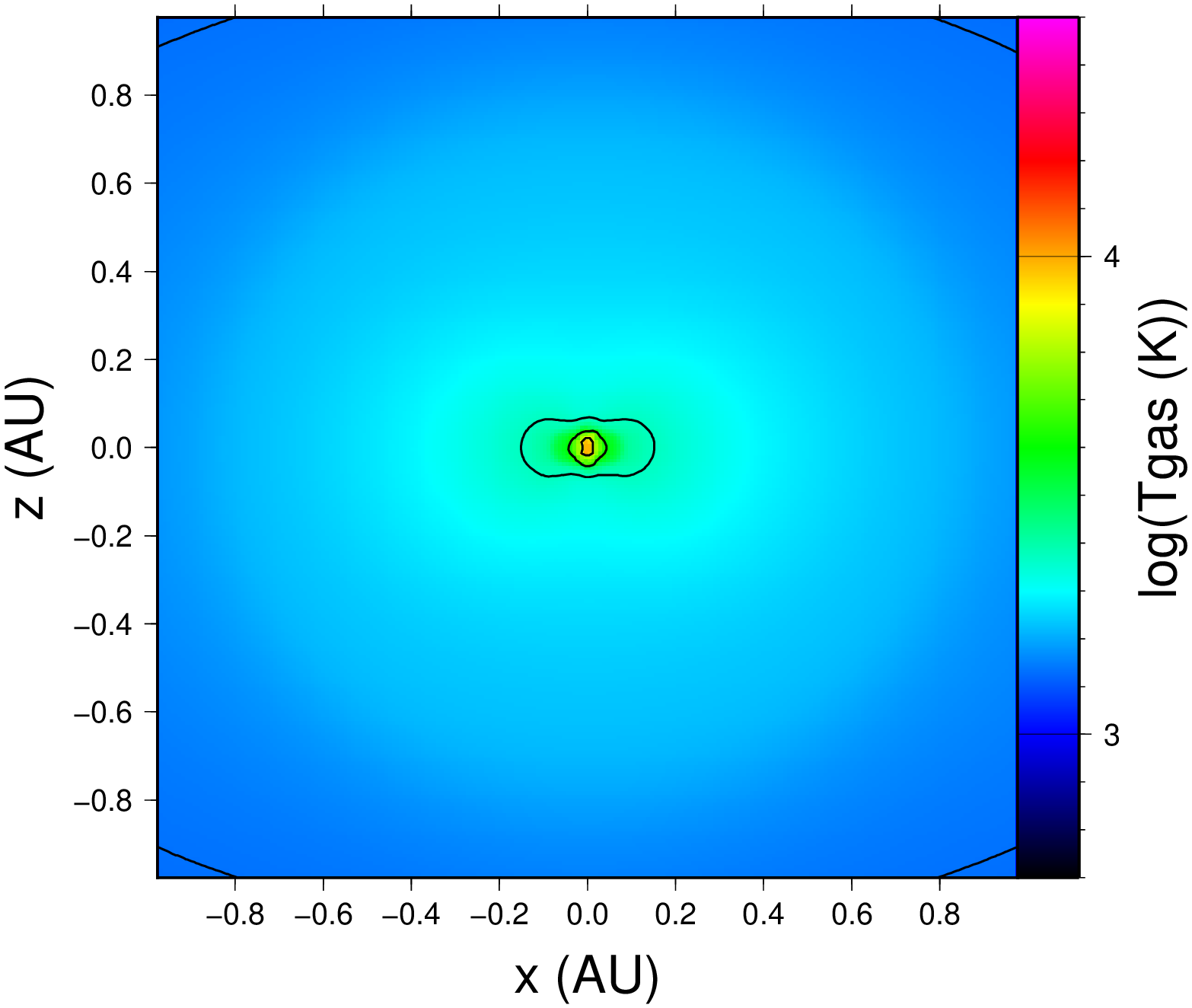}
\includegraphics[width=40mm,trim=23mm 0mm 10mm 0mm,clip]{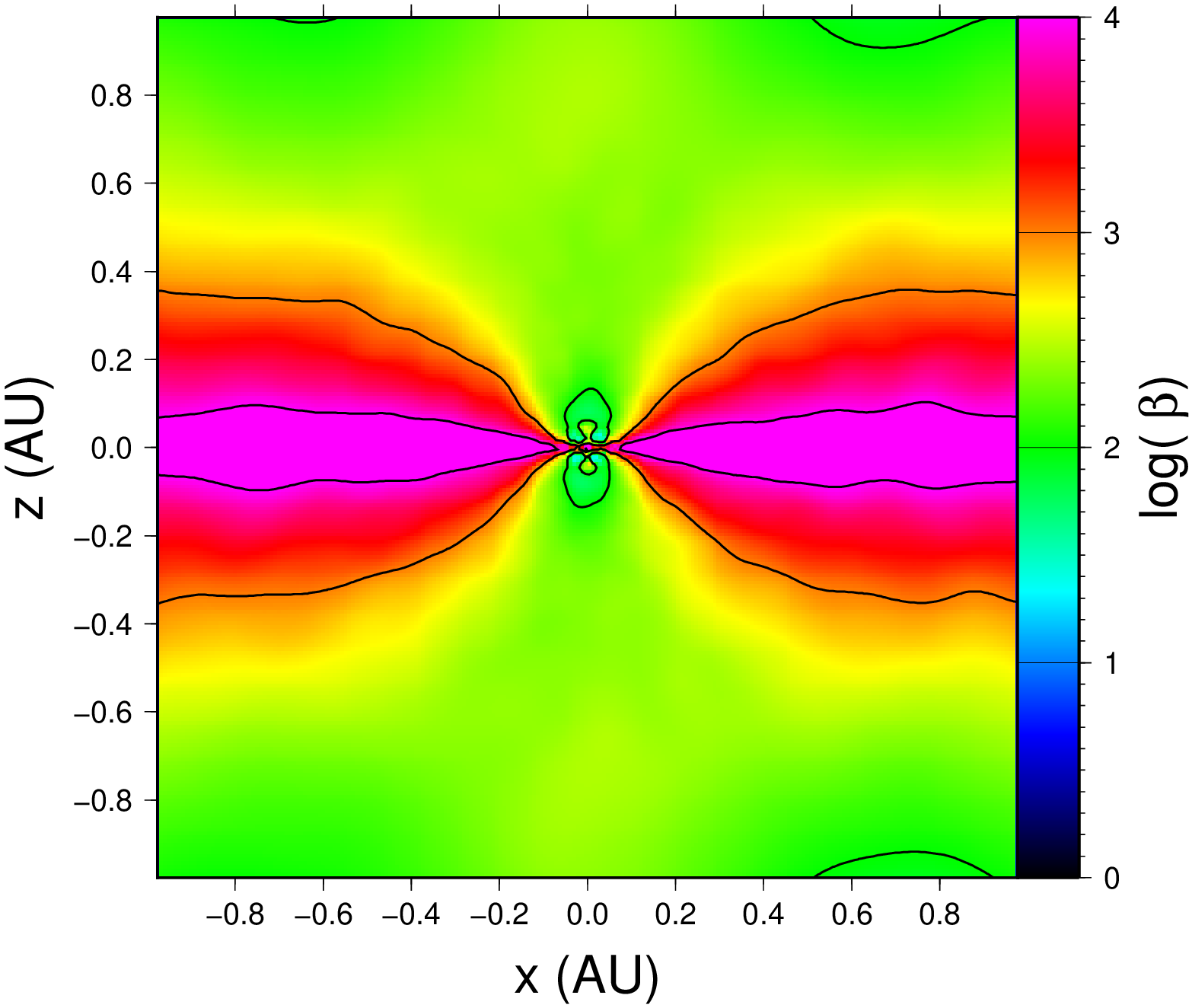}
\caption{
The cross sections of the density, gas temperature, and plasma $\beta$ (from left to right) around the protostar in the $y=0$ plane.
The top, middle, and bottom row corresponds to model 1, 2, and 3, respectively.
The thin black lines show the contours of each quantity.
The color bars of the density, temperature, and plasma $\beta$ 
are expressed as 
$\log(\rho ~[\cm])$, $\log(T ~[{\rm K}]) $, and $\log(\beta)$, respectively.
Note that the x, y, and color-bar scales differ between 
the ideal model and the resistive models.
}
\label{2dmap_around_secondcore}
\end{figure*}

When the central density reaches 
$\rho_c\sim 10^{-3} \cm$ and the hydrogen molecules are completely 
dissociated, the gas evolves adiabatically and 
the protostar forms at the centre of the first core.
In figure \ref{2dmap_around_secondcore}, 
we show the cross sections of density, 
temperature, and plasma $\beta$ around the protostar. 
The central density 
is  $\rho_c\sim 10^{-3} \cm$ at this epoch and just 
after the protostar formation.
Note that the x, y, and color-bar scales differ between the ideal model and
resistive models because the structure around the 
protostar in the ideal model is
quantitatively different from the one in the other models.
The density distributions of the resistive 
models (middle and bottom left panels) exhibit the dumbbell-like structures.
These structures indicate that the rotation plays a 
role in the resistive models.
On the other hand, in model 1 (the ideal model), the density structure is 
elliptical and there is no dumbbell-like structure 
even in vicinity of the protostar.
As we will show below,
the rotationally supported disc quickly 
forms during the subsequent evolution in the resistive models 
but does not form in the ideal model.
The temperature distributions around the protostar are smooth and
roughly spherically symmetric in all models.
The temperature exceeds $1000$ K and the magnetic diffusion is no longer
effective in the entire region.
In the model 2, the low $\beta$ region forms in the 
vertical direction. This structure is created by the rotational 
amplification of the magnetic field.
As a result, the plasma $\beta$ becomes 
$\beta \sim 10^{-1}$.
The magnetic field is also magnified in model 3. However,
it is not a significant magnification 
and the plasma $\beta$ in the vertical direction is
still $\beta\sim 10^2$ at this epoch.
We cannot find any signature of the rotational amplification 
in  model 1. 
The low $\beta$ region in the vertical direction is created by a
dragging of the poloidal magnetic field. 
The figure \ref{2dmap_around_secondcore}  shows 
that the structures around the protostar are significantly
different even just after the protostar formation when 
the magnetic diffusion is considered.



\begin{figure*}
\includegraphics[width=40mm,angle=-90]{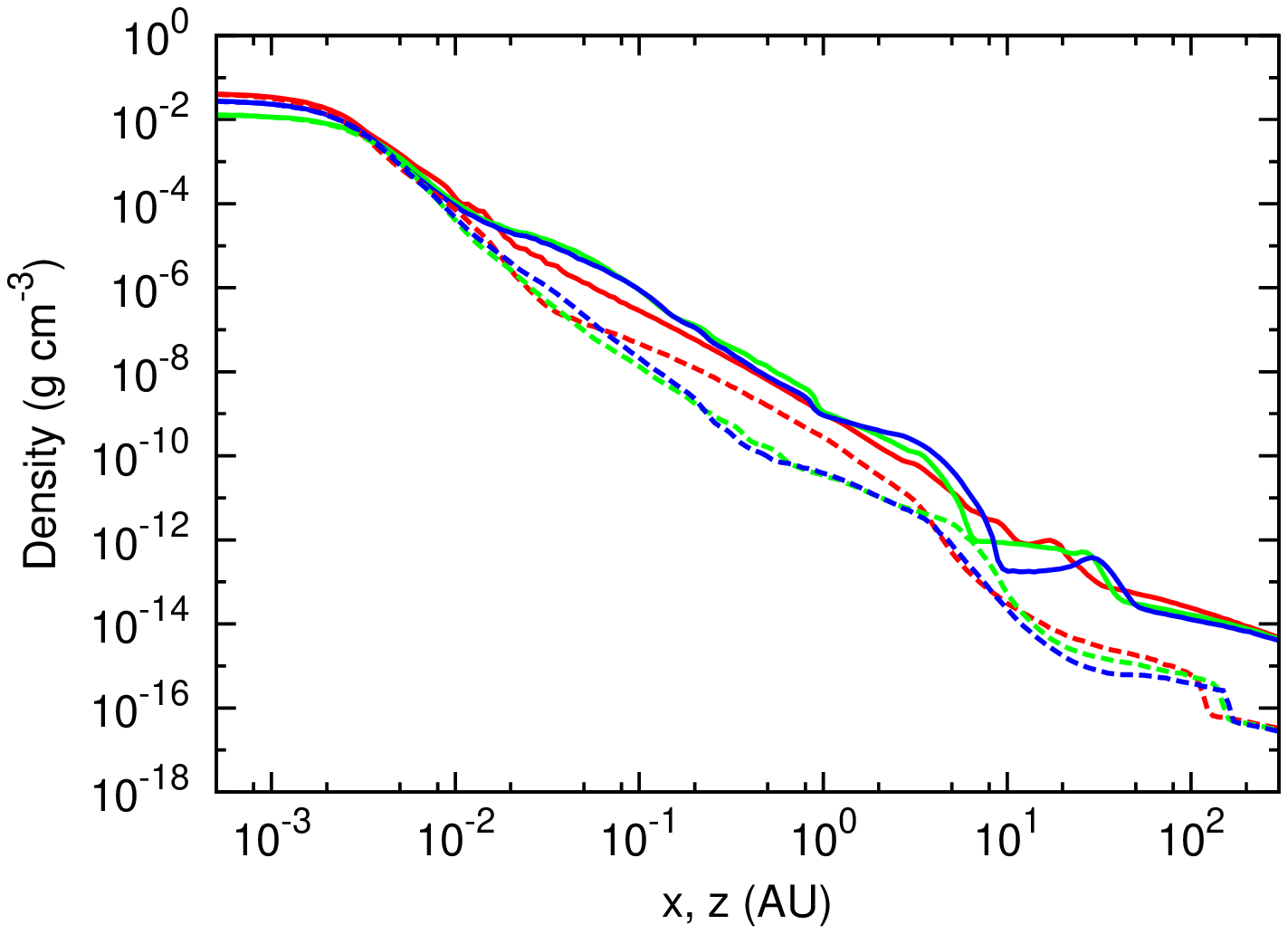}
\includegraphics[width=40mm,angle=-90]{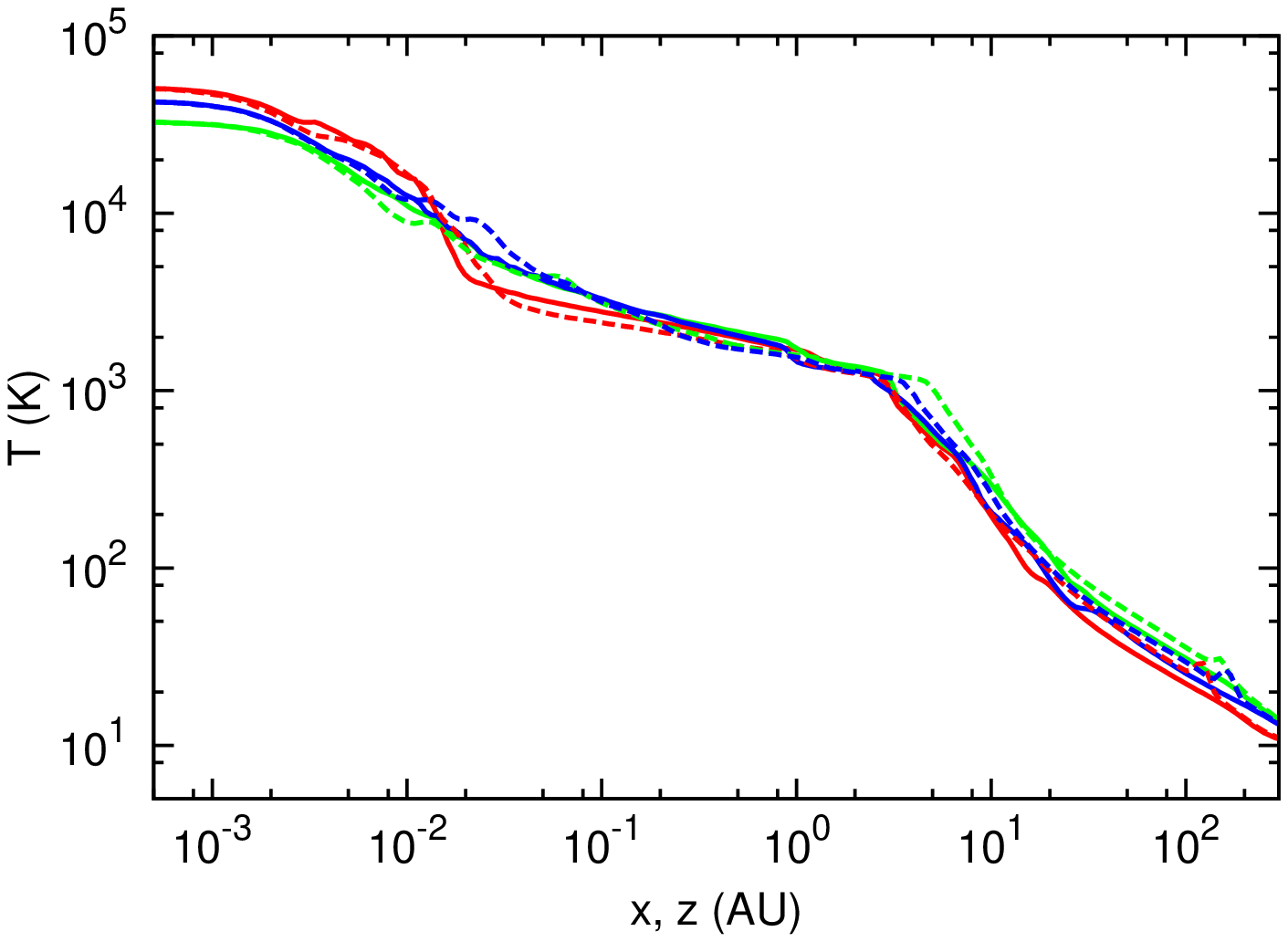}
\caption{
The profiles of the density (left) and gas temperature (right) at the end of
the simulations.
The solid and dashed lines show the profiles in the $x$ and $z$ directions, respectively.
The red, green, and blue lines show the results of model 1, 2, and 3, respectively.
}
\label{rho_and_Tgas}
\end{figure*}

After the protostar forms,
it evolves via the mass accretion 
from the remnant of the first core. 
In figure \ref{rho_and_Tgas}, we show the density and gas temperature along the
$x$-axis (solid lines) and $z$-axis (dashed lines) at the end of the simulations.
The central densities and temperatures reach 
$\rho_c\sim 10^{-2}-10^{-1} \cm$ and $T_c \gtrsim 10^4 $ K, respectively. 
From the decrease in the density and temperature of the red lines
around $x,z\sim 10^{-2} {\rm AU}$,
we can identify the radius of the protostar in the ideal model as
$r \sim 10^{-2} {\rm AU}$.
In the ideal model, the difference between the density in the horizontal 
and the vertical directions 
is not large and the density structure is almost spherically symmetric.
On the other hand, the density profiles of the resistive models show a 
different structure around the protostar.
After the formation of the protostar, 
the rotationally supported disc of size $1$ AU quickly
forms in resistive models in these epochs.
Because of the disc formation, the boundary of the protostar 
becomes ambiguous in the density and temperature profiles 
in the horizontal direction.
Weak shock wave structures can be seen at $x \sim 1 $ AU in the
green and blue solid lines of density.
This is the boundary of the circumstellar discs.

\begin{figure*}
\includegraphics[width=40mm,angle=-90]{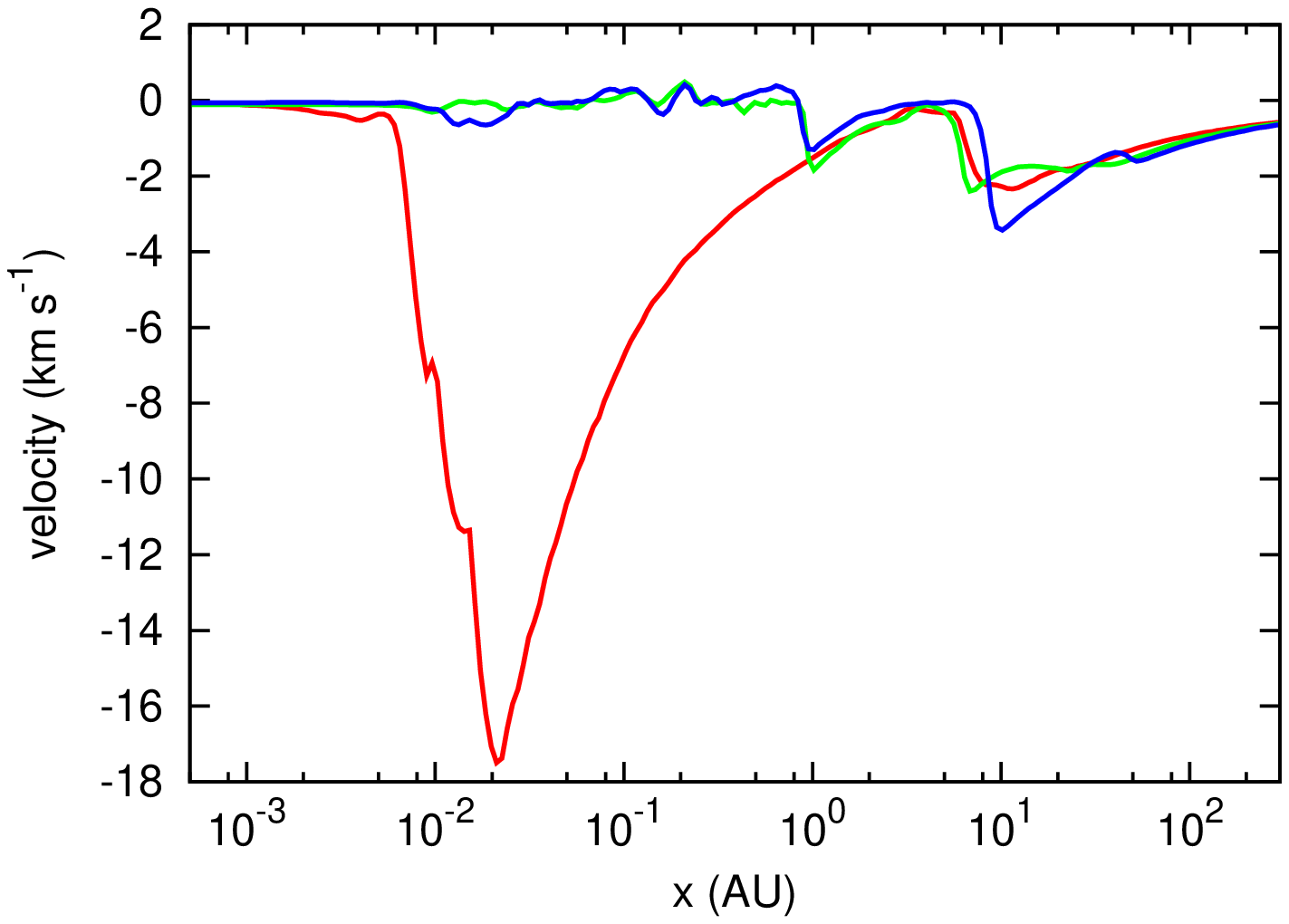}
\includegraphics[width=40mm,angle=-90]{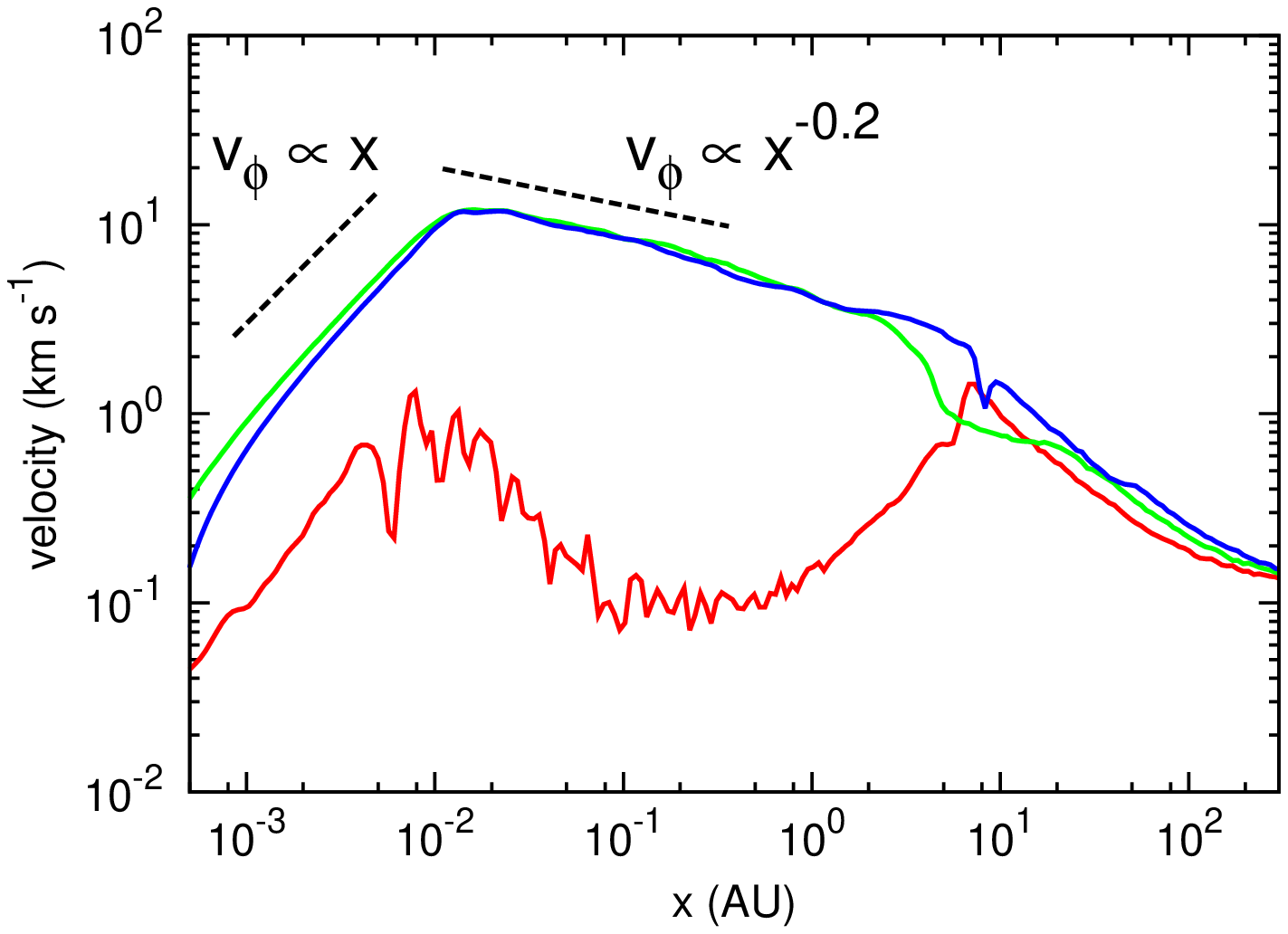}
\caption{
The profiles of the infall velocity (left) and rotation velocity (right) in the $x$ direction.
The epochs are the same as in figure \ref{rho_and_Tgas}.
The red, green, and blue lines are also defined as in figure \ref{rho_and_Tgas}.
}
\label{vx_and_vy}
\end{figure*}

In figure \ref{vx_and_vy}, we show the infall 
and rotation velocity along the $x$-axis.
The left panel shows the infall velocity. 
In the ideal model, the infall reaches $x\sim 10^{-2}$ AU, which
shows that the first core remnant
accretes directly onto the central protostar. 
On the other hand, the infall stops
at $x \sim 1 $ AU in the resistive models. 
This radius corresponds to the shocks in the density profiles
and thus to the edges of the discs.
Note that there are the other shocks at $x\sim10$ AU. These are
the accretion shocks at the surface of the first core. 
The remnant of the first core still exists in these epochs.


We can see a clear 
transition of the rotation profile at $x \sim 10^{-2}$ AU
in the resistive models (blue and green lines).
In $x\lesssim 10^{-2}$ AU, the profile obeys 
$v_\phi \propto x$ and the gas rigidly rotates.
This rigidly rotating region is the protostar and 
its radius in the resistive models
is also $r\sim 10^{-2}$ AU. 
In $10^{-2}\lesssim x\lesssim 1 $ AU, 
the profile follows $v_\phi \propto x^{-0.2}$.
This is the rotation profile of the disc around the protostar. 
The rotation profile of the disc is more shallow than 
for a Keplerian disc (or disc subjected to a
gravitational potential created by a point mass) 
which obeys the profile of $v_\phi \propto x^{-0.5}$.
This means that both the self-gravity of the disc and the gravity of 
the central protostar influence the rotation profile.

\subsection{Rotationally supported disc around protostar}
As we have seen above, there are several features of the density 
and velocity profiles which suggest the existence of a circumstellar disc.
For example, the rotational velocity at the mid-plane of
models 2 and 3 is considerably larger than the radial velocity
in $10^{-2}\lesssim x\lesssim 1$ AU.
In addition, shocks exist at $x\sim 1$ AU in the density and
infall velocity profiles.
However, it is not clear from the above analysis whether the 
disc is rotationally supported or not.

\begin{figure}
\includegraphics[width=60mm,angle=-90]{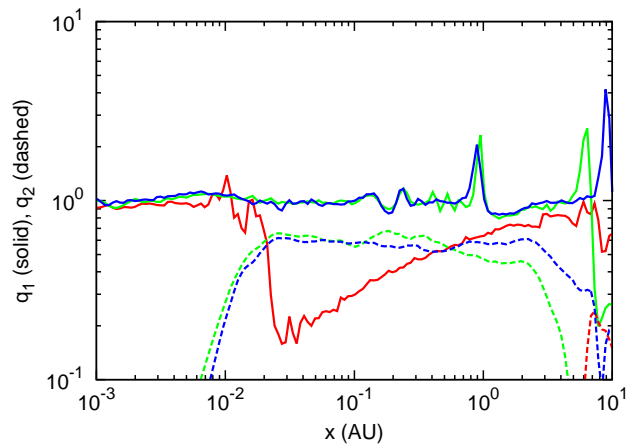}
\caption{
Solid lines show the ratio of the sum of the centrifugal 
force and the pressure gradient force
to the radial gravitational force,
$
q_1=|\frac{v_\phi^2/r+\nabla_r p/\rho}{\nabla_r \Phi}|,
$
as a function of the radius.
Here, $p$ and $\Phi$ are the pressure and the gravitational potential, respectively.
The dashed lines show the ratio of the centrifugal force to the
radial gravitational force,
$
q_2=|\frac{v_\phi^2/r}{\nabla_r \Phi}|.
$
The red, green, and blue lines show the results of models 1, 2, 
and 3, respectively.
The epochs are the same as in figure \ref{rho_and_Tgas}.
}
\label{ratio_rot}
\end{figure}

To confirm that the disc is really rotationally supported,
the ratio of the sum of the centrifugal 
and the pressure gradient forces
to the radial gravitational force,
\begin{equation}
q_1=|\frac{v_\phi^2/r+\nabla_r p/\rho}{\nabla_r \Phi}|,
\end{equation}
is plotted in figure \ref{ratio_rot}
with the solid lines
and the ratio of the centrifugal to the
radial gravitational force,
\begin{equation}
q_2=|\frac{v_\phi^2/r}{\nabla_r \Phi}|,
\end{equation}
with the dashed lines.
Here, $p$ and $\Phi$ are the pressure 
and the gravitational potential, respectively.
When $q_1=1$ and $q_2 \ll q_1$, 
the gas is supported by the pressure gradient force.
On the other hand, when $q_1=1$ and $q_2 \sim q_1$, 
the gas is mainly supported by the centrifugal force.

The red lines show that $q_1 \sim 1$ and $q_2 \ll q_1$ 
for $x \lesssim 10^{-2} {\rm AU}$. This means that a
pressure supported second core (the protostar), whose radius
is $r\sim 10^{-2}  {\rm AU}$ exists at the centre. On the other hand,
the radial gravitational force always dominates other forces 
for $10^{-2} {\rm AU} \lesssim x \lesssim 5 {\rm AU}$. Therefore, 
neither the pressure gradient force nor the centrifugal force can cancel 
the gravitational collapse and no rotationally supported disc forms in
the ideal model. On the other hand, 
the green and blue lines
show that $q_1$ is almost unity for $x \lesssim 1 {\rm AU}$ and
the gravitational force is cancelled in this region.
For $x \lesssim 10^{-2} {\rm AU}$, the $q_1 \sim 1$ and $q_2 \ll q_1$,
which shows the existence of a pressure supported protostar.
Meanwhile, $q_2$ is about $ 0.6$ for 
$10^{-2} {\rm AU}\lesssim x \lesssim 1 {\rm AU}$
and 60\% of
the gravitational force is cancelled by the centrifugal
force and the remaining 40\% is cancelled by the
pressure gradient force in this region.
Thus, the gas is supported mainly by the centrifugal force.
From these results, we conclude that the rotationally supported
disc forms naturally  in the very early phase of the protostar formation
when the magnetic resistivity is included and the first core phase
is considered correctly.
Note that the dips of the green and blue solid lines at the edge of the 
disc are due to the large pressure gradient there. 
The ram pressure caused by the mass accretion should balance this.

\begin{figure}
\includegraphics[width=60mm,angle=-90]{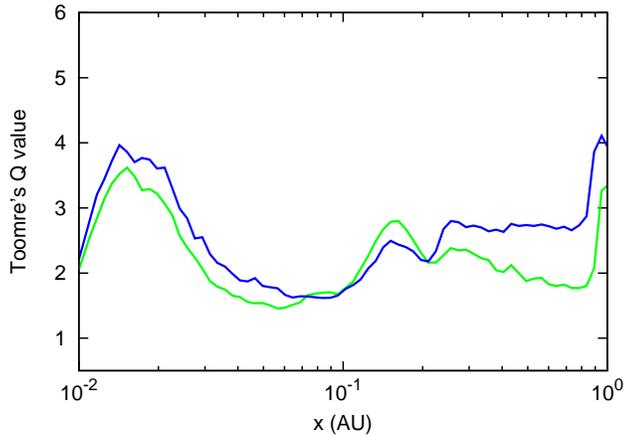}
\caption{
Toomre's $Q$ value as a function of the radius in the $x$ direction.
The green and blue lines show the results of model 2 and 3, respectively.
The epochs are the same as in figure \ref{rho_and_Tgas}.
}
\label{Qvalue}
\end{figure}

The first core directly becomes the disc and its mass 
is much larger than that of the protostar during its formation epoch.
Thus, it is expected that the self gravity plays an important role in the
early phase of the disc evolution \citep{2010ApJ...718L..58I}.
In figure \ref{Qvalue}, we show Toomre's $Q$ value of 
the disc $Q=\frac{c_s \Omega}{\pi G \Sigma}$,
where we approximate the epicycle frequency $\kappa$ as $\Omega$.
In the disc region
$10^{-2} {\rm AU} \lesssim x \lesssim 1 {\rm AU}$ 
Toomre's Q value is $Q \sim 2-3$. 
As pointed out  in previous studies,
the disc becomes unstable against non-axisymmetric perturbations 
when $Q\sim 1.5$ and the spiral arms develop  \citep{1998ApJ...504..945L}. 
The spiral arms invoke an angular 
momentum transfer.
Although, the Q value is still slightly larger than 1.5,
it is expected that the gravitational instability plays a very 
important role for the angular momentum transfer in the 
subsequent disc evolution because a large amount of the 
remnant of the first core is still accreating to the disc
and the disc mass increases quickly.

\section{Summary and Discussions}

In this paper, we investigated the formation and evolution of 
the first core, the protostar and the disc around the protostar by
using three-dimensional simulations with radiation transfer, 
as well as Ohmic and ambipolar diffusions. 

Our findings are summarized as follows.
\begin{enumerate}
\item The magnetic flux is largely removed 
in the first core phase.
As a result, at the centre of the first core,  
plasma $\beta$ becomes $\beta>10^4$. On the other hand, the $\beta$ at the
centre of the first core in ideal simulation is $\beta\sim 10$.

\item Even though the plasma $\beta$ inside the first core is significantly
different in the resistive and the ideal models, the angular momentum
of the first core is not (within an order of 
magnitude). This is because most of the angular momentum has been  
removed before the magnetic diffusion processes play a role. 
Actually, figure 11 of \citet{2007ApJ...670.1198M} suggests 
that most of the angular momentum is removed from the gas
during the isothermal collapse phase.
When the magnetic field is neglected, 
a disc with $r\sim 100$ AU forms in the cloud core 
for the parameters adopted in our simulations
\citep[see, e.g.,][]{2011MNRAS.416..591T,2015MNRAS.446.1175T}.
This also suggests that most of the angular momentum is removed during
the isothermal phase.

\item With magnetic diffusions, a circumstellar 
disc forms around the protostar just after protostar formation
 even with a relatively strong initial magnetic field 
(we employ a uniform density sphere and
an initial mass-to-flux ratio relative to the critical value of $\mu=4$). 
We confirmed that the disc is rotationally supported.
The disc is massive enough to enable gravitational instability to develop in the
subsequent disc evolution. Thus, the gravitational instability plays an
important role in the early evolution of the circumstellar discs.

\end{enumerate}

The reason 
why most of the angular momentum is removed from the gas in the 
isothermal collapse phase can be understood
by comparing the magnetic braking timescale 
$t_{\rm b} \sim \lambda_{\rm J}/v_{\rm A}$ to 
the free-fall timescale $t_{\rm ff}$,
where $\lambda_{\rm J}$ and $v_{\rm A}$ are the Jeans length 
and Alfv\'{e}n velocity, respectively.
The magnetic braking timescale is estimated as the time in which the 
inertia of the central region is equal to the inertia of the envelope where
the Alfv\'{e}n wave sweeps \citep{2004ApJ...616..266M}.
The ratio of the two timescale $t_{\rm b}/t_{\rm ff}$ 
is given as 
$t_{\rm b}/t_{\rm ff}\sim \lambda_{\rm J}/(v_{\rm A} t_{\rm ff})\sim \sqrt{\beta}$.
In our simulations, the plasma $\beta$ is $\beta=1.7$ at the 
initial condition ($\rho=5.5\times 10^{-18} \cm$) and decreases during the 
early isothermal collapse phase
as $\beta\propto c_s^2/v_{\rm A}^2 \propto \rho^{-1/3}$, where we assume
that $c_s$ is constant and $B\propto \rho^{2/3}$ 
as shown in figure \ref{rho_B}. When the central density
reaches $\rho_c=10^{-15} \cm$, $t_{\rm b}/t_{\rm ff}=\sqrt{\beta}=0.71$ and
the magnetic braking timescale becomes shorter than the 
free fall timescale.
Therefore, $t_{\rm b}/t_{\rm ff} \lesssim 1$ 
and the angular momentum is largely removed 
during the isothermal collapse phase.

Our results about the disc formation are largely consistent with
those of the previous studies which followed the protostar formation with
sufficient resolution and considered
the first core phase \citep[e.g.,][]{
2011MNRAS.413.2767M,2013ApJ...763....6T}.
We believe that the development of a disc at the 
very early phase of the star formation is a robust consequence.
The previous research we mentioned above considered only Ohmic diffusion.
On the other hand, we also included ambipolar diffusion.
This does not change the overall
formation process of the disc significantly.
However, it is possible that the ambipolar diffusion plays a
more important role in the subsequent evolution of the disc because 
it extends the density range in which
the magnetic field and the gas are decoupled and allows the magnetic flux
to escape from the disc.

The difference in disc formation between the ideal model and resistive models
is due to the strength of the magnetic field
and not the difference in the angular momentum of the first core.
In our simulations, the circumstellar disc forms in the resistive models
(model 2 and 3) and does not in the ideal model (model 1).
As we have seen above, in resistive models, 
the plasma $\beta$ of the envelope around the protostar is
$\beta \gtrsim 10^1$ except for the vicinity of the protostar of model 2
(the middle and bottom right panels of figure \ref{2dmap_around_secondcore})
and the magnetic braking is ineffective.
On the other hand, the magnetic field removes the angular momentum from the gas 
during the second collapse in the ideal model because the plasma $\beta$ of 
envelope is $10^{-1}<\beta<10^{1}$ 
(see, the top right panel of figure \ref{2dmap_around_secondcore})
and the magnetic braking timescale is comparable or less than
the free-fall timescale ($t_{\rm b}/t_{\rm ff}\sim \sqrt{\beta}$).
This is why the circumstellar disc does not form in the ideal model.
The simulation with Ohmic diffusion in \citet{2015ApJ...801..117T}  showed that 
the circumstellar disc forms even in the slowly rotating first core 
($J\sim 2 \times 10^{50}~{\rm g~cm^2~s^{-1}}$ where $J$ is the 
angular momentum). Thus, the several-fold difference in the angular momentum 
does not affect whether or not the disc forms.

Because the magnetic flux is largely removed in the first core phase,
the proper treatment of the first core is 
necessary to investigate the formation of the protostar and disc.
In previous works which argue that the disc formation is 
strongly suppressed by the magnetic braking \citep[e.g.,][]
{2008ApJ...681.1356M,2011ApJ...738..180L}, 
the inner boundary was set from the beginning of the simulations.
With this treatment,
the previous works cannot follow the first core phase properly that
should be supported by gas pressure.
The discrepancy between our results and those of these works
should be due to the different treatments of the first core 
phase \citep[see, also][]{2012A&A...541A..35D}.

It is expected that the disc size becomes larger 
than the size obtained in our simulations ($r<1$ AU) once
the mass accretion from the remnant of the first core finishes 
because the massive remnant still exists and is accreating onto the disc,
even at the end of the simulations. 
Unfortunately, it is almost impossible to investigate 
the further evolution of the disc without a sink.
Although the sink may introduce  numerical 
artefacts (especially in the few sink radius), it
is an essential technique for investigating the 
long-term evolution of the disc.
We will investigate the further evolution of the disc with 
the sink technique while
remembering that this introduces numerical artefacts.

In this paper, we showed that the SPH method is capable of 
treating MHD and non-ideal processes in realistic astrophysical simulations.
Our results are largely consistent with those of the recent non-ideal
RMHD simulations with the static-mesh-refinement 
code \citep{2015ApJ...801..117T}.
Thus, our method is reliable and 
can be used for astrophysical simulations.
Because the SPH method is relatively easily implemented and more flexible
than static-mesh-refinement code, it can be used as an alternative method for 
many astrophysical problems in which the magnetic field 
play the important role.

In the simulations presented in this paper, several 
approximations were adopted.
The influences of these simplifications 
should be investigated in future studies.
For example, we used a fixed dust grain size of $a=0.035~ {\rm \mu m}$ 
and a fixed the cosmic-ray ionization of $\xi_{\rm CR}=10^{-17}~ {\rm s^{-1}}$.
The latter is not good approximation 
for the dense region, $\rho \sim 10^{-11} \cm$.
We also used a simple rigidly rotating gas sphere as the initial condition.
As \citet{2012A&A...543A.128J} and \citet{2014MNRAS.438.2278M} have pointed 
out, the initial density profile and the magnetic field configuration
strongly affect the size of the circumstellar discs.
In future, we will investigate how the differences in the initial configuration
affect the disc evolution.




\section *{Acknowledgments}
We thank  Dr. K. Tomida,  Dr. T. Matsumoto, 
and Dr. D. Stamatellos for their fruitful discussions.
We also thank Dr. K. Tomida and Dr. Y. Hori to provide their EOS table to us.
We also thank anonymous referee for helpful comments.
The computations were performed on a parallel 
computer, XC30 system at CfCA of NAOJ.
Y.T. and K.I are financially supported by Research 
Fellowships of JSPS for Young Scientists.

\bibliography{article}

\end{document}